\documentclass[fleqn,usenatbib]{mnras}

\usepackage{newtxtext,newtxmath}

\usepackage[T1]{fontenc}

\DeclareRobustCommand{\VAN}[3]{#2}
\let\VANthebibliography\thebibliography
\def\thebibliography{\DeclareRobustCommand{\VAN}[3]{##3}\VANthebibliography}

\usepackage{graphicx}	
\usepackage{amsmath}	
\usepackage{nccmath}
\usepackage{multirow,booktabs,colortbl,tabularx}

\def\lcdm{$\Lambda$CDM }

\def\gassf{gas$_{\rm{SF}}$ }

\definecolor{Mygrey}{gray}{0.75}

\title[BH growth in misaligned EAGLE galaxies]{Stellar-gas kinematic misalignments in \textsc{eagle}: enhanced SMBH growth in misaligned galaxies}

\author[Maximilian K. Baker et al.]{
Maximilian K. Baker$^{1}$\thanks{E-mail: bakermk@cardiff.ac.uk},
Timothy A. Davis$^{1}$,
Freeke van de Voort$^{1}$,
Sandra I. Raimundo$^{2,3}$
\\
$^{1}$Cardiff Hub for Astrophysics Research \&\ Technology, School of Physics \&\ Astronomy, Cardiff University, Queens Buildings, Cardiff, CF24 3AA, UK
\\
$^{2}$DARK, Niels Bohr Institute, University of Copenhagen, Jagtvej
155, Copenhagen N, 2200, Denmark
\\
$^{3}$Department of Physics \&\ Astronomy, University of Southampton, Highfield, Southampton, SO17 1BJ, UK
}
\date{Accepted XXX. Received YYY; in original form ZZZ}

\pubyear{\the\year{}}

\begin{document}
\label{firstpage}
\pagerange{\pageref{firstpage}--\pageref{lastpage}}
\maketitle

\begin{abstract}
Stellar-gas kinematic misalignments are a transient phenomenon observed in $\sim11$~per~cent of the local galaxy population. According to current models, misaligned gas is expected to lose angular momentum and relax into the galactic plane on timescales of $\sim0.1$~Gyr, driving gas toward the central regions of the galaxy. Recent observational studies have found a higher incidence of active galactic nuclei in misaligned galaxies. We use the \textsc{eagle} simulation to explore the connection between stellar-gas misalignments and enhanced central black hole (BH) activity between $0<z<1$. We use a sample of $\sim5600$ galaxies with a stellar mass of $M_{*}\geqslant \mathrm{10^{9.5}}$~M$_\odot$ that feature long-lived stellar-gas alignment, counter-rotation, and unstable misalignments (non-coplanarity). Over time windows of $0.5$~Gyr, we find that galaxies experiencing an unstable misalignment have systematically enhanced BH growth during relaxation. Galaxies with long-term counter-rotation show little difference in BH growth compared to aligned galaxies. We suggest that this enhanced BH growth is driven by loss of angular momentum in unstable misaligned gas discs which is able to drive gas inward toward the vicinity of the BH. At $z=0.1$, we find a greater incidence of overmassive BHs in galaxies that have spent a greater fraction of time with unstable stellar-gas kinematic misalignments over the preceding $\approx2$~Gyr compared to control samples of aligned galaxies. In agreement with observations, we conclude that BH activity is enhanced in misaligned systems in \textsc{eagle} and suggest that the presence of overmassive BHs may be indicative of a past stellar-gas kinematic misalignment.

\end{abstract}




\section{Introduction}\label{intro:}

Supermassive black holes (SMBHs) are predicted to exist at the centre of most massive galaxies. Observational studies have found strong correlations between the properties of the central SMBH and the host galaxy \citep[e.g.][]{SilkRees98, FerrareseMerritt00, MarconiHunt03}. For example, a strong linear correlation has been found between the mass of the SMBH and the mass of the galactic bulge \citep[e.g.][]{HaringRix04, KormendyHo13}. Moreover, positive correlations have been found between the SMBH accretion rate and galactic star formation rate (SFR) \citep[see][and references therein]{HeckmanBest14}. At high BH accretion rates, the SMBH becomes visible as an active galactic nucleus (AGN). The energy released by an AGN is thought to be able to affect their host galaxies and is a key requirement in many theoretical models to explain the quenching observed in many local red-sequence early-type galaxies (ETGs). These relations suggest that galaxies co-evolve with their SMBH \citep[see ][for a review]{KormendyHo13} and that the supply of cold, molecular gas within galaxies is intricately linked to episodes of SF and AGN activity \citep[e.g.][]{Shlosman90, StorchiSchnorr2019}.

Stellar-gas kinematic misalignments (henceforth `misaligned galaxies') have become a key probe in understanding cold gas replenishment in galaxies \citep[e.g.][]{Sarzi2006, Davis2011, DavisBureau2016, Bryant2019, Duckworth2020a, Baker2024}. Kinematic misalignments are observed in $\approx11$~per~cent of local galaxies \citep{Bryant2019, Raimundo2023}, including $\sim30-40$~per~cent of ETGs and $\sim5$~per~cent of late-type galaxies \citep[LTGs; e.g.][]{Davis2011, Bryant2019, Ristea2022, Raimundo2023}. Gas-rich minor mergers were frequently proposed as the dominant formation pathway for misaligned galaxies among early observational studies \citep[e.g.][]{Davis2011}. However, recent results from hydrodynamical simulations have suggested a more diverse formation pathway with contributions from internal gas replenishment mechanisms \citep[e.g. cooling of hot halo gas; ][]{Keres05, Lagos2014}, external gas replenishment mechanisms \citep[e.g. gas-rich mergers; ][]{Khim2021, Baker2024}, alongside pertubations of the existing co-rotating gas induced from fly-bys or ram-pressure stripping \citep[e.g.][]{Starkenburg2019, Khim2021, Casanueva2022}. 

As outlined in \citet{Tohline1982}, any kinematically misaligned component is subject to a radially-dependent torque from the mass distribution within the galactic plane \citep[see also][]{LakeNorman1983}. For a misaligned gas disc, neighbouring rings of gas become non-coplanar and dissipate angular momentum through cloud-cloud collisions \citep[][]{VoortDavis2015}. Interactions between co-rotating gas from stellar mass loss \citep[e.g.][]{ParriottBregman08, LeitnerKravtsov11} and counter-rotating gas act as a further source of dissipation \citep[e.g.][]{Negri2014, CapeloDotti2017, Taylor2018, Khoperskov2020, Peirani2025}. Consequently, the misaligned gas disc is able to relax into the galactic plane, while driving gas inward \citep[][]{VoortDavis2015}. In the absence of ongoing smooth accretion, this relaxation process is predicted to be short-lived with timescales of $\sim100$~Myr for a typical ETG \citep[][]{DavisBureau2016}. Using the \textsc{eagle} simulation \citep[][]{SchayeCrain2015, Crain2015calibration}, \citet{Baker2024} found that the majority ($\approx80$~per~cent) of misalignments relaxes within the expected timescales, with median relaxation timescales of $\sim0.5$~Gyr. However, longer relaxations ($\gtrsim2$~Gyr) far exceeding theoretical predictions have also been found \citep[][]{VoortDavis2015, Khim2021, Baker2024}. Therefore, kinematic misalignments may provide a mechanism for transporting cold gas into the inner sub-kpc regions of the galaxy for enhanced SF and AGN activity \citep[][]{VoortDavis2015, Khoperskov2020, Duckworth2020b}.

Correlations between the presence of misalignments and enhanced central SF \citep[e.g.][]{Chen2016, Xu2022} or AGN activity \citep[e.g.][]{Penny2018, Raimundo2021, Ristea2022, Raimundo2023, Raimundo2025, Winiarska2025} have been found in several observational studies of nearby galaxies with the use of Integral Field Surveys (IFU). For instance, kinematic mapping of radio galaxies NGC 3100 \citep{Ruffa2019b}, MCG-06-30-15 \citep{Raimundo2013} and NGC 5077 \citep{Raimundo2021} reveal the presence of misaligned gas discs, with the resulting inflows acting as a likely driver of AGN activity. Initial studies aiming to establish a connection between AGN activity and misalignments were often limited to low sample sizes \citep[e.g.][]{Penny2018, Ilha2019}. However, a significant correlation was recently established by \citet{Raimundo2023} using a sample of $\sim1300$ galaxies from the Sydney-AAO Multi-object Integral field spectrograph (SAMI) galaxy survey \citep[][]{Croom12, Bryant15}. AGNs were found to preferentially reside in misaligned galaxies ($\approx17$~per~cent) compared to aligned galaxies ($\approx7$~per~cent) at the $3\sigma$ confidence level. Furthermore, the higher incidence of AGNs found among ETGs ($\approx58$~per~cent) compared to LTGs ($\approx42$~per~cent) was attributed to the higher incidence (and longevity) of misalignments in ETGs. As such, \citet{Raimundo2023} highlight the importance of stellar-gas misalignments, in addition to the presence of gas, as a mechanism to trigger AGN activity in galaxies \citep[see also][]{Raimundo2025}.  

One of the challenges with observations is the difficulty in discerning whether the observed AGN signatures in misaligned galaxies is driven by the misaligned gas disc itself or whether these are a 'smoking gun' from the initial gas accretion that also formed the misalignment \citep[see][]{Raimundo2025}. This is because misaligned external accretion (e.g. minor mergers) has been shown both as an effective formation pathway for misaligned discs \citep[e.g.][]{VoortDavis2015, Khim2021, Baker2024} and, more directly, as a means of driving gas inward through interactions with the in-situ gas and triggering central SF or AGN activity \citep[][]{Sales2012, Kaviraj2014c, Taylor2018, DavisYoung2019, Ellison2024c, Peirani2025}. Additionally, outflows driven by episodes of AGN activity, especially in low-mass systems \citep[][]{Penny2018, Starkenburg2019}, may act as a source of co-rotating gas depletion and disrupt the orientation of inflowing gas \citep[][]{Ristea2022, Casanueva2022, Cenci2023}. This may increase the likelihood of misalignment formation, especially if the outflowing material is injected into a misaligned halo \citep[e.g.][]{Lagos2015, Duckworth2020a}. As such, observed AGN signatures in misaligned galaxies are often non-trivial to interpret.

Cosmological hydrodynamical simulations such as \textsc{eagle} and IllustrisTNG \citep[][]{Nelson19} have become useful tools to understand the behavior of misaligned galaxies in the context of gas inflows and AGN fueling. For instance, accreting misaligned or counter-rotating gas has been established as an effective means to drive gas inward and promote SF through dissipative interactions with in-situ, co-rotating gas \citep[e.g.][]{ThakarRyden1996, Sales2012, Taylor2018, Starkenburg2019, Davies2022, Han2024}. This mechanism has also been found to trigger associated AGN activity \citep[e.g.][]{Taylor2018, Khoperskov2020, Duckworth2020b}. Likewise, misaligned galaxies have been linked with a lower specific gas angular momentum \citep[][]{Starkenburg2019, Duckworth2020a} and more compact SF regions \citep[e.g.][]{Casanueva2022}. Using the IllustrisTNG simulation, \citet{Duckworth2020b} find that misaligned low-mass galaxies have higher peak BH luminosities and enhanced BH growth (as a proxy for integrated BH activity) over the past 8~Gyr compared to an aligned control sample. These results suggest the formation and persistence of a stellar-gas kinematic misalignment may act as a key source of gas inflow and AGN activity in addition to any merger-driven activity.

In this paper, we use the \textsc{eagle} (Evolution and Assembly of GaLaxies and their Environments) suite of cosmological hydrodynamical simulations to investigate the connection between AGN activity and ongoing misalignments. We use the sample of relaxing stellar-gas kinematic misalignments between $0<z<1$ from \citet{Baker2024}. Specifically, we aim to investigate whether: 1) BH growth is enhanced in galaxies that are currently unstably misaligned (i.e. their gas discs are relaxing into the dynamically stable co- or counter-rotating regime) compared to samples of stable (aligned and counter-rotating) systems, and 2) the degree to which this imprinted on the galaxy population in the form of overmassive SMBHs at low redshifts.

This paper is structured as follows. In Section~\ref{EAGLE:} we give an overview of the \textsc{eagle} simulations. In Section~\ref{Sample:} we outline the sample of galaxies we use for this work. In Section~\ref{Methods:} we give an overview of our methods. In Section~\ref{results:} we present our results. In Section~\ref{discussion:} we discuss and compare our results to existing results. Finally, we conclude in Section~\ref{conclusion:}.


\section{The EAGLE simulation}\label{EAGLE:} 

\textsc{eagle}\footnote{Publicly available data products available from \url{http://eagle.strw.leidenuniv.nl} and \url{https://icc.dur.ac.uk/Eagle} and described in \citet{Mcalpine2016}} is a suite of cosmological hydrodynamical simulations described in \citet{SchayeCrain2015} and \citet{Crain2015calibration}. The \textsc{eagle} simulations have been successful in reproducing the observed stellar mass, size, and morphology distributions \citep[e.g.,][]{SchayeCrain2015, Correa2017, TrayfordSchaye2018, Hill2021}. These simulations are run using a heavily modified version of \textsc{Gadget}-3 \citep{Springel05, Springel2008}, with an SPH formulation known as \textsc{Anarchy} (see \citealt{Schaller2015} for description). Galaxy evolution can be traced using the `subhalo' merger trees by \citet{Qu2017} that were generated using the \textsc{D-Trees} algorithm \citep{Jiang14}. A \lcdm cosmology is assumed with $\Omega _{\rm{m}} = 0.307$, $\Omega _{\rm{\Lambda}} = 0.693$, $\Omega _{\rm{b}} = 0.04825$, $h = 0.6777$, $\sigma _{\rm{8}} = 0.8288$, $n_{\rm{s}} = 0.9611$ and $Y = 0.248$ \citep{PlanckCollaboration14}.

\textsc{eagle} relies on sub-grid modelling in order to model physical processes below the resolution limit. These include radiative cooling and photoheating \citep{Wiersma09a}, star formation \citep{SchayeDallaVecchia08}, stellar evolution and interstellar medium (ISM) enrichment \citep{Wiersma09b}, stochastic thermal stellar feedback \citep{DallaVecchiaSchaye12}, and BH accretion and feedback outlined below.

Black hole (BH) particles are seeded with a mass of $1.48\times10^{5}$~M$_{\odot}$ in the centres of dark matter haloes that have a minimum total mass of $1.48\times10^{10}$~M$_{\odot}$ and that do not already contain a black hole particle, following the method of \citet{Springel05}. The sub-grid BH mass, $M_{\mathrm{BH}}$, is allowed to grow via a modified Bondi-Hoyle accretion prescription that accounts for the velocity of surrounding gas, as described in detail by \citet{RosasGuevara15}. This sub-grid accretion rate, $\Dot{M}_{\mathrm{BH}}$, is then capped at the Eddington accretion rate. Upon accretion of mass, energy is stored in a reservoir that is injected stochastically into surrounding gas particles using single-mode thermal AGN feedback, as described in \citet{BoothSchaye09}. The sub-grid parameters in \textsc{eagle} are tuned to reproduce observations of the galaxy stellar mass function and galaxy sizes relation at $z\approx0$ \citep[for details, see ][]{Crain2015calibration}.

For this work, we use the largest volume simulation of the reference model known as \textsc{Ref-L100N1504}. This simulation has a maximum physical gravitational softening length of 0.7 pkpc (proper kiloparsec), a box-length of 100 cMpc (co-moving megaparsec), containing an initially equal number ($1504^3$) of dark matter and baryonic particles, with initial particle masses of $9.7\times10^6$~M$_\odot$ and $1.81\times10^6$~M$_\odot$, respectively. The outputs for these data are stored on 400 "snipshots", of which we use 200 (with a mean time cadence of $\approx120$~Myr within $0<z<1$) for which merger trees were run by \citet[][]{Crain2017}.


\section{Sample}\label{Sample:}

We sample galaxies within $0<z<1$ using the same selection criteria as \citet{Baker2024}, with additional considerations made for reliable BH data. Here we summarise the selection criteria. For full details, see \citet{Baker2024}:

\begin{enumerate}
 \item A minimum stellar mass of $M_{*}>10^{9.5}$~M$_{\odot}$ within a spherical aperture size of 30~pkpc. 
 \item A minimum star-forming gas (gas$_{\rm{SF}}$) particle count of 20 within the stellar half-mass radius, $r_{50}$, which is defined as the spherical radius enclosing half the stellar mass within 30 pkpc of the stellar centre of mass.
 \item A maximum spatial separation of $<2$~pkpc between the stellar and gas$_{\rm{SF}}$ centres of mass within $r_{50}$.
 \item A maximum $1\sigma$ uncertainty of $<30^{\circ}$ for the misalignment angle using bootstrap resampling. Misalignment angles, $\psi_{\mathrm{3D}}$, are measured in three-dimensional (3D) space between the stellar and gas$_{\rm{SF}}$ angular momentum vectors within $r_{50}$.
\end{enumerate}

As in \citet{Baker2024}, we note that our results and conclusions are unchanged if we impose a higher particle limit of 100 gas$_{\rm{SF}}$ particles, or require a $3\sigma$ uncertainty of $<30^{\circ}$. 

As we intend to investigate the intrinsic behaviour of misaligned gas in the absence of bias from the viewing angle, we focus on misalignment angles in 3D space \citep[though these are not significantly different to projected 2D angles, see e.g.][]{Baker2024}. We classify galaxies into three classifications; aligned galaxies with angles $\psi_{\mathrm{3D}}<30^{\circ}$, \textit{unstable} misaligned galaxies with angles $30^{\circ} \leq \psi_{\mathrm{3D}} \leq 150^{\circ}$, and counter-rotating galaxies with angles $\psi_{\mathrm{3D}}>150^{\circ}$, following widely used classifications \citep[e.g.][]{Davis2011}. 

For this work, we focus on the change in BH mass as a proxy for cumulative BH activity. This is because we are limited by the low temporal resolution of our snipshots ($\sim10^8$~yr) relative to the timescales of AGN flickering \citep[$\sim10^{5}$~yr; ][]{Schawinski15}. As highlighted by \citet{Crain2015calibration}, gas accretion for BH particles near the seed mass ($\sim1.48\times10^{5}$~M$_{\odot}$) tends to be slow, because Bondi accretion rates scale with $M_{\mathrm{BH}}^2$. Likewise, mergers between low-mass BHs in this regime constitute a significant fraction of the BH's mass, manifesting as characteristic BH masses at multiples of the BH seed mass. As such, we focus our analysis on BHs with a minimum mass of $M_{\mathrm{BH}}>10^{6}$~M$_{\odot}$ using the most massive BH particle within $r_{50}$. Using the BH closest to the galactic center of potential leaves our results unchanged. 

In order to compare BH growth between galaxies in varying stages of kinematic stability, we trace the BH growth over $0.50\pm0.05$~Gyr windows in galaxies that meet our selection criteria. This window-size is chosen in order to probe long-term BH growth in galaxies that spend a significant amount of time ($\sim0.5$~Gyr) with an unstable misaligned gas disc. Furthermore, we require no BH mass decreases over this time window and no spontaneous increase in BH mass by a factor of $>5$ between consecutive snipshots. These criteria ensure that we trace the same BH particle over the $0.5$~Gyr window, while ensuring that we do not attribute significant BH growth if a secondary, lower-mass BH particle is erroneously identified as the main BH over one or more snipshots.

Aligned and counter-rotating samples are created by selecting random $0.50\pm0.05$~Gyr windows in which $\psi_{\mathrm{3D}}<30^{\circ}$ and $\psi_{\mathrm{3D}}>150^{\circ}$ for the full $0.50\pm0.05$~Gyr period, respectively. For our sample of unstable misalignments, we utilise the parent sample of 3154 misalignment relaxations used and explained in detail in \citet{Baker2024}. In short, this is a sample of galaxies that experience an unstable misalignment ($30^{\circ}\leq\psi_{\mathrm{3D}}\leq150^{\circ}$) and are traced while the gas disc is kinematically unstable (defined as $20^{\circ}\leq\psi_{\mathrm{3D}}\leq160^{\circ}$). We note that the first snipshot in the unstable regime typically coincides with the first snipshot in the unstable \textit{misaligned} regime. An unstable disc is considered `settled' upon returning to the kinematically stable regime (defined as $\psi_{\mathrm{3D}}<20^{\circ}$ and $\psi_{\mathrm{3D}}>160^{\circ}$), so long as the galaxy remains in the stable regime for a period of at least another $0.1$~Gyr. As in \citet{Baker2024}, we define the relaxation time, $t_{\rm{relax}}$, between the first snipshot in the kinematically unstable regime, and the first snipshot in which the unstable disc has settled. In order to fit within the target window duration of $0.50\pm0.05$~Gyr, we extract a sample of relaxations with $t_{\rm{relax}}\geq 0.45$~Gyr. 

This leaves us with a total sample of 5570 galaxies, of which 4657 are continuously aligned, 532 are continuously counter-rotating, and 381 form our `unstable misaligned' sample that are in the process of relaxing back into the stable regime (i.e. the galactic plane). Each of these samples has a median temporal window of $\approx0.51$~Gyr.

\section{Methodology}\label{Methods:}

As in \citet{Baker2024}, we make use of the co-rotational energy fraction as a proxy for morphology for stellar and \gassf components \citep[see][]{Correa2017}. This describes the fraction of the total kinetic energy ($K$) invested in ordered co-rotation ($K^{\rm{rot}}_{\rm{co}}$) and is given by 
\begin{ceqn}
    \begin{align}
    \kappa _{\rm{co}} = \frac{K_{\rm{co}}^{\rm{rot}}}{K} = \frac{1}{K}\sum_{i, L_{\rm{z}, \textit{i}} > 0} \frac{1}{2}m_{i}\left ( \frac{L_{\rm{z}, \textit{i}}}{m_i R_i} \right )^{2},
    \end{align}
\end{ceqn}
where the sum is over all particles of a given type (stars or gas$_{\mathrm{SF}}$) within a spherical radius ($30$~pkpc for stars and $r_{50}$ for gas$_{\mathrm{SF}}$) centred on the minimum potential, $m_{i}$ is the particle mass, $L_{\rm{z},\textit{i}}$ is the particle angular momentum component along the direction of the total angular momentum of a given particle type of the galaxy, and $R_{i}$ is the radius from the centre of potential in the plane normal to the rotation axis of the given particle type \citep{Correa2017}. The stellar and \gassf co-rotational energy fractions are denoted as $\kappa_{\rm{co}}^*$ and $\kappa_{\rm{co}}^{\rm{SF}}$, respectively.

As shown in \citet{Correa2017}, a value of $\kappa_{\rm{co}}^*=0.4$ can be used to approximately distinguish the `red sequence' quiescent spheroidal galaxies (ETGs with $\kappa_{\rm{co}}^*<0.4$) from the `blue cloud' discy star-forming galaxies (LTGs with $\kappa_{\rm{co}}^*>0.4$). Likewise, as shown in \citet{Jimenez2023}, $\kappa_{\rm{co}}^{\rm{SF}}\gtrsim0.7$ approximately corresponds to a thin gas disc which we confirmed by visual inspection. We note that during the initial stages of misalignment formation and the resulting dissipative effects, we expect $\kappa_{\rm{co}}^{\rm{SF}}$ to be naturally lower in unstable misaligned galaxies than in dynamically relaxed galaxies. 

We define the star-forming gas fraction as 
\begin{ceqn}
    \begin{align}
 f_{\rm{gas,SF}} = \frac{M_{\rm{gas,SF}}}{M_{\rm{gas}}+M_{*}},
    \end{align}
\end{ceqn}
where $M_{*}$, $M_{\rm{gas}}$, and $M_{\rm{gas,SF}}$ are the stellar, gas, and gas$_{\rm{SF}}$ masses of the galaxy within the kinematic aperture of $r_{50}$ akin to \citet{Baker2024}.

In order to gain a rough idea of the incidence of AGN within each subsample, we can use the sub-grid BH accretion rate to estimate the bolometric luminosity and Eddington ratio of the BH. A common expression \citep[e.g.][]{Habouzit22} to convert the BH accretion rate in cosmological simulations to a bolometric luminosity is given by
\begin{ceqn}
    \begin{align}
 L_{\rm{bol}} = \frac{\epsilon_{r}}{1-\epsilon_{r}} \dot{M}_{\mathrm{BH}} c^{2},
    \end{align}
\end{ceqn}
where $\epsilon_{r}=0.1$ is the radiative efficiency and $c$ is the speed of light in a vacuum. 

Additionally, we define the Eddington ratio ($\lambda_{\mathrm{Edd}}$) as the ratio between the current sub-grid BH accretion rate and the theoretically maximum spherically-symmetric Eddington accretion rate given by 
\begin{ceqn}
    \begin{align}
 \dot{M}_{\mathrm{Edd}} = \frac{4\pi G M_{\mathrm{BH}} m_{p}}{\epsilon_{r}\sigma_{T}c},
    \end{align}
\end{ceqn}
where $G$ is the gravitational constant, $m_{p}$ is the proton mass, and $\sigma_{T}$ is the Thomson scattering cross-section for an electron. 


\section{Results}\label{results:} 

\subsection{BH growth}\label{results: enhanced BH growth}

\subsubsection{Total galaxy population}\label{results: enhanced BH growth total}

\begin{figure}
    \begin{center}
    \includegraphics[width=8.5cm]{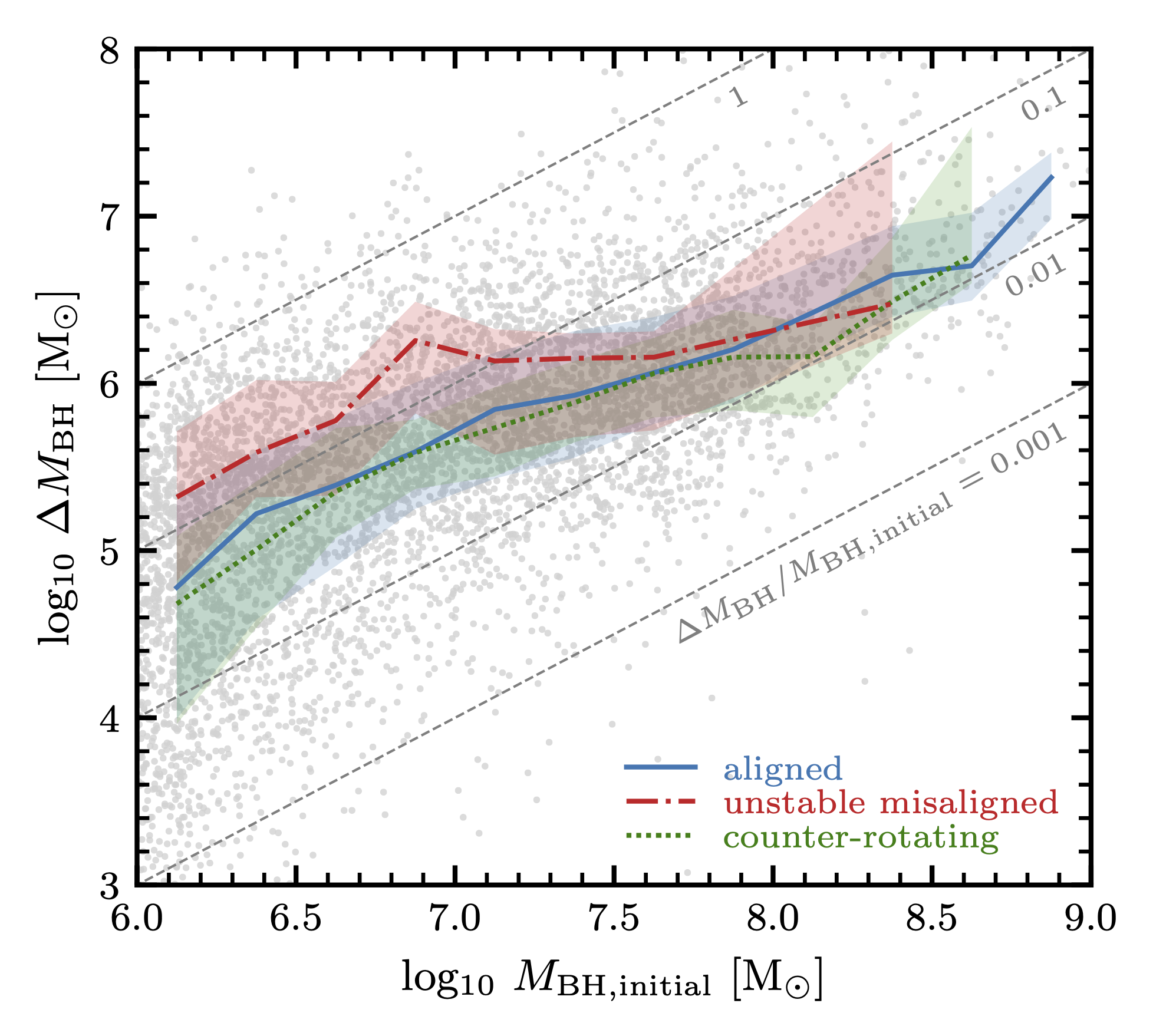}\\
    \caption{Median change in BH mass traced over a $0.50\pm0.05$~Gyr window as a function of the initial BH mass for samples of galaxies which are aligned (blue solid), counter-rotating (green dotted), and galaxies experiencing an unstable misalignment (red dash-dotted) for bins with a minimum bin count of 10. Shaded regions show the $25$ and $75$ percentiles. Grey scatter points show the total sample. Diagonal lines (dashed grey) show the relative increase in BH mass, with $\Delta M_{\mathrm{BH}}/ M_{\mathrm{BH,initial}}=1$ indicating a doubling of BH mass over this window. We find that counter-rotating galaxies exhibit similar BH growth to aligned galaxies, while galaxies experiencing an unstable misalignment show significantly higher ($\sim0.6$~dex) BH growth up to BH masses of $M_{\mathrm{BH, initial}} \lesssim 10^{7.5}$~M$_{\odot}$.}
    \label{plot: BH growth scatter}
    \end{center}
\end{figure}

\begin{figure}
    \begin{center}
    \includegraphics[width=8.5cm]{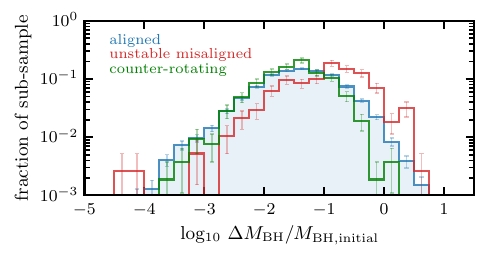}\\
    \caption{Fractional BH growth distributions over a $0.50\pm0.05$~Gyr window for sub-samples of aligned (blue), counter-rotating (green), and galaxies experiencing an unstable misalignment (red). Errors are given as Poisson uncertainties. We find that unstable misaligned galaxies have higher fractional BH growth with a median of $0.129\Delta M_{\mathrm{BH}}/ M_{\mathrm{BH, initial}}$, while aligned and counter-rotating galaxies have similar growth fractions of $0.039\Delta M_{\mathrm{BH}}/ M_{\mathrm{BH, initial}}$ and $0.034\Delta M_{\mathrm{BH}}/ M_{\mathrm{BH, initial}}$, respectively.}
    \label{plot: BH growth hist}
    \end{center}
\end{figure}

In Figure~\ref{plot: BH growth scatter} we show the growth of the BH mass ($\Delta M_{\mathrm{BH}}$) over our $\sim0.5$~Gyr windows compared to the initial BH mass at the start of the window ($M_{\mathrm{BH, initial}}$). These results are also shown in Figure~\ref{plot: BH growth hist}, in which the BH growth is expressed as a fractional increase in BH mass from the initial BH mass, with distributions given for sub-samples of aligned, unstable misaligned, and counter-rotating galaxies.

As seen clearly in Figure~\ref{plot: BH growth scatter}, we find galaxies relaxing from an unstable misalignment experience significantly enhanced ($\sim0.6$~dex) BH growth compared to samples of continuously aligned and counter-rotating galaxies. This is especially true at lower masses of $M_{\mathrm{BH, initial}} \lesssim 10^{7.4}$~M$_{\odot}$. At higher BH masses of $M_{\mathrm{BH, initial}} \gtrsim 10^{7.5}$~M$_{\odot}$, unstable misaligned galaxies show comparable BH growth to their aligned and counter-rotating counterparts. However, we caution that the sample size of unstable misaligned galaxies with $M_{\mathrm{BH, initial}} \gtrsim 10^{7.5}$~M$_{\odot}$ is low. Over the entire range of BH masses considered, unstable misaligned galaxies grow their BHs by $\approx12.9$~per~cent on average over $0.5$~Gyr, while aligned systems grew by only $\approx3.9$~per~cent. In order to test the significance of this result, we perform a two-sample Kolmogorov–Smirnov (KS) test on the distributions in Figure~\ref{plot: BH growth hist}. We obtain a KS-test statistic $=0.32$ and p-value $\ll 10^{-10}$, indicating a strongly-significant result given the size of the sample. 

Interestingly, counter-rotating systems share a similar BH growth to aligned galaxies within the full BH mass range considered (see Figure~\ref{plot: BH growth scatter}). Counter-rotating galaxies grow their BHs by a median of $3.4$~per~cent compared to $3.9$~per~cent for aligned galaxies (KS-test statistic $=0.12$, p-value $=2.0\times10^{-6}$). We discuss possible interpretations for this result in Section~\ref{results: gas reservoirs and dynamics}.

These results remain unchanged if we use the instantaneous sub-grid BH accretion rate, $\dot{M}_{\mathrm{BH}}$, time-averaged over the same $\approx0.5$~Gyr window (not shown). We note that the use of the sub-grid BH accretion rate should be taken with caution given the significant fluctuations seen in $\dot{M}_{\mathrm{BH}}$ that may be poorly captured within the snipshot time cadence of $\approx120$~Myr \citep[e.g.][]{McAlpine2017}. Among BHs with $M_{\mathrm{BH, initial}}\sim 10^{6.5}$~M$_{\odot}$, both aligned and counter-rotating galaxies have BH accretion rates of $\Dot{M}_{\mathrm{BH}}\sim1\times10^{-4}$~M$_{\odot}$~yr$^{-1}$, averaged over $\approx0.5$~Gyr. Galaxies experiencing an unstable misalignment within this mass range have enhanced sub-grid BH accretion rates of $\Dot{M}_{\mathrm{BH}}\sim4\times10^{-4}$~M$_{\odot}$~yr$^{-1}$. Alternatively, we can estimate the average accretion rate required to grow the BH by $\Delta M_{\mathrm{BH}}$ between each consecutive snipshot. Taking the radiative efficiency as $\epsilon_{r}=0.1$, we find an average BH accretion rate of $\sim7\times10^{-4}$~M$_{\odot}$~yr$^{-1}$ among unstable misaligned galaxies with $M_{\mathrm{BH, initial}}\sim 10^{6.5}$~M$_{\odot}$.

Evaluating the peak value of $\dot{M}_{\mathrm{BH}}$ attained over our $\approx0.5$~Gyr window for sub-samples of aligned/unstable misaligned/counter-rotating galaxies, we find $\approx46/50/43$~per~cent of BHs reaching bolometric luminosities of $L_{\rm{bol}}>10^{43}$~erg~s$^{-1}$, and $\approx37/54/28$~per~cent of BHs with Eddington ratios of $\lambda_{\mathrm{Edd}}>0.01$. These results are consistent with the observation that AGN signatures are more likely to be observed in misaligned systems \citep[][]{Raimundo2023, Raimundo2025}.

By inspection of individual BH evolutions, we find that the BH accretion rate tends to be enhanced for the duration of the unstable misalignment rather than an initial single snipshot of high accretion. We also tested using the median values of the BH accretion rate, obtained from either the subgrid BH accretion values or $\Delta M_{\mathrm{BH}}$, instead of averages and found our results did not change. Furthermore, misaligned galaxies in our $\approx0.75$~Gyr and $\approx1.0$~Gyr samples also clearly show this enhanced BH growth.

\subsubsection{Early-type and late-type morphologies}\label{results: enhanced BH growth ETGs and LTGs}

\begin{figure}
    \begin{center}
    \includegraphics[width=8.5cm]{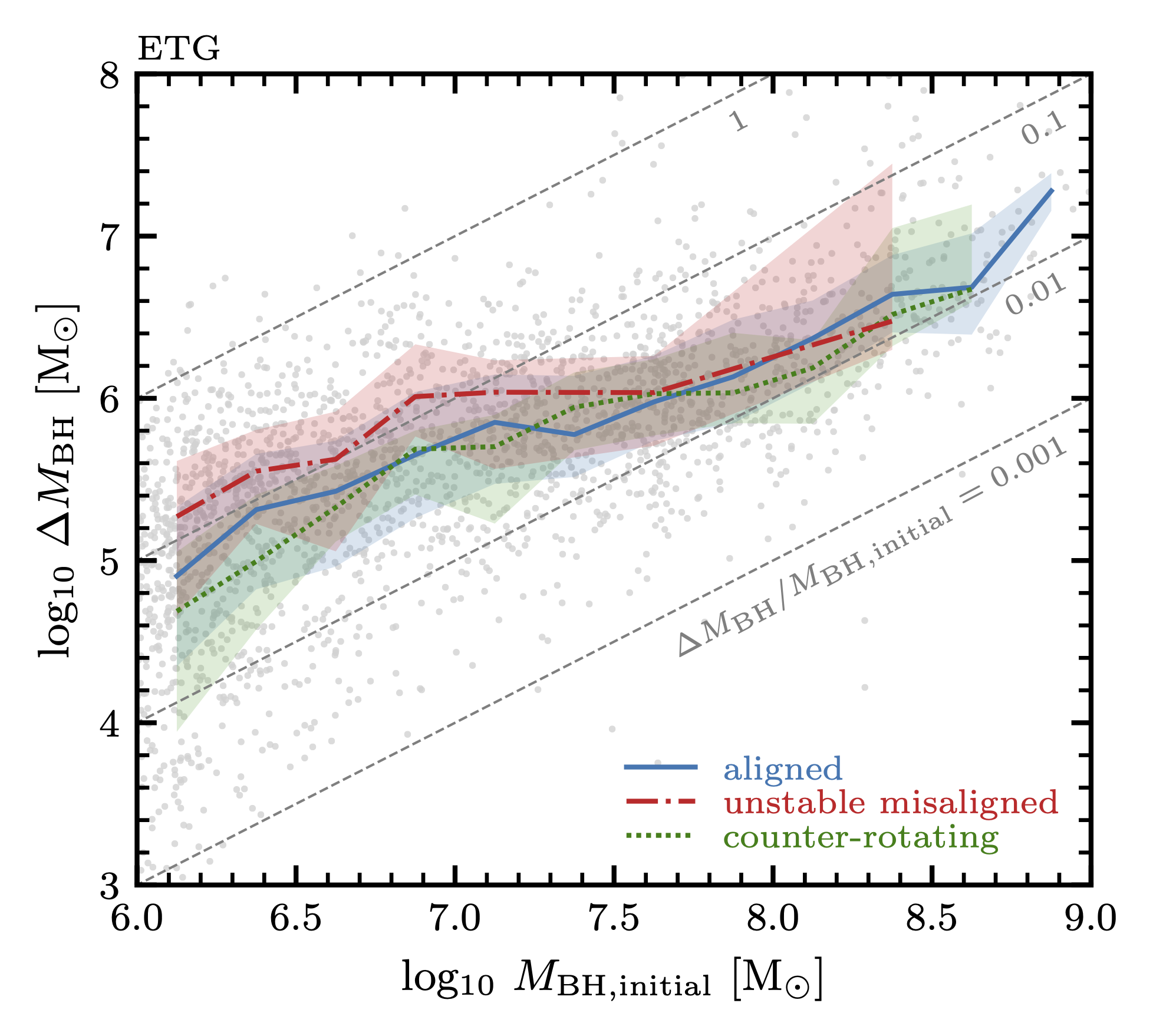}\\
    \includegraphics[width=8.5cm]{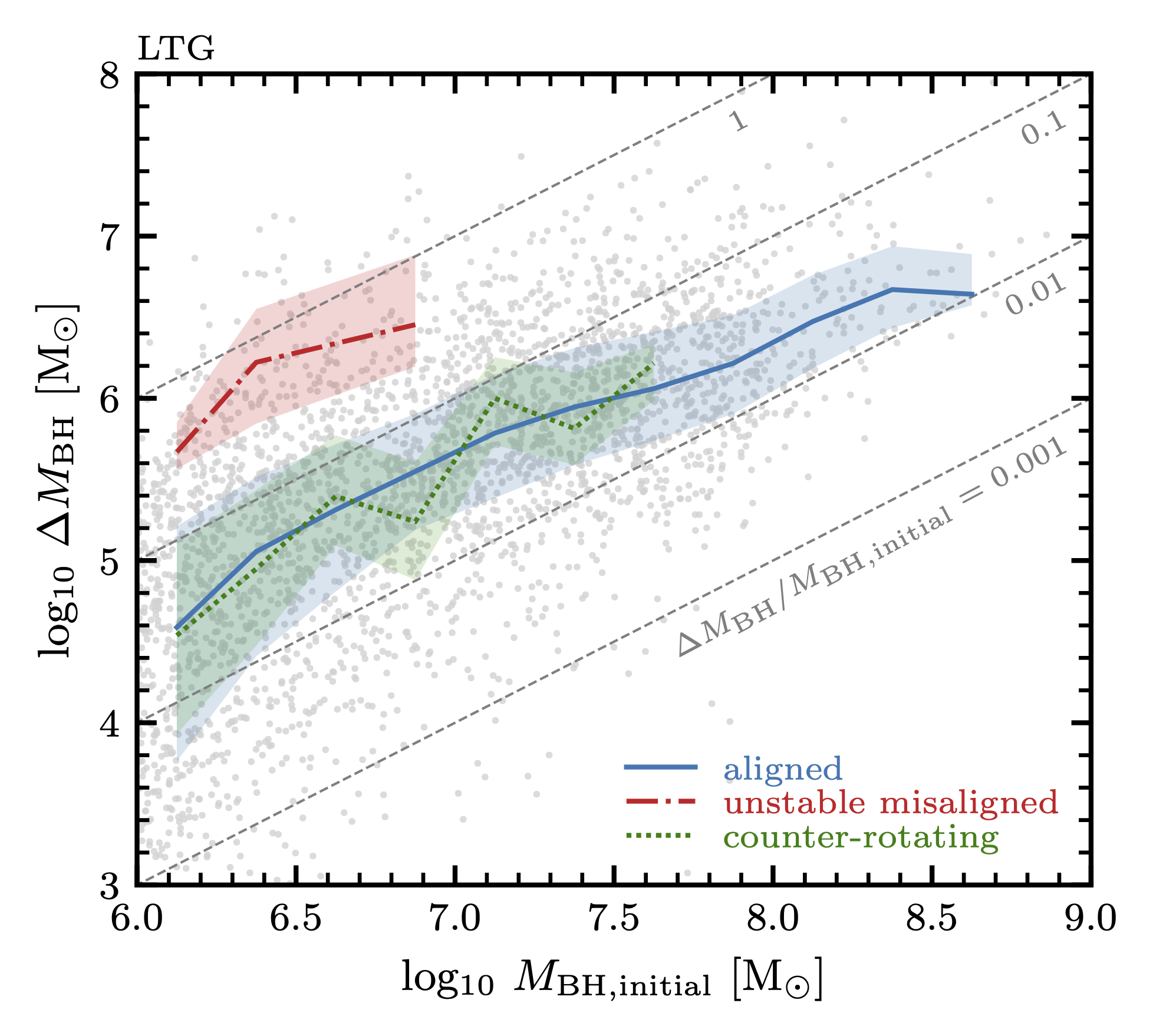}\\
    \caption{Median change in BH mass traced over a $0.50\pm0.05$~Gyr window as a function of the initial BH mass for samples of ETGs (top) and LTGs (bottom) which are aligned (blue solid), counter-rotating (green dotted), and galaxies experiencing an unstable misalignment (red dash-dotted) for bins with at least 10 galaxies. Shaded regions show the $25$ and $75$ percentiles. Grey scatter points show the total sample. Diagonal lines (dashed grey) show the relative increase in BH mass, with $\Delta M_{\mathrm{BH}}/ M_{\mathrm{BH,initial}}=1$ indicating a doubling of BH mass over this window. We find that BH growth is strongly enhanced ($\sim1.0$~dex) in unstable misaligned LTGs compared to aligned and counter-rotating morphological counterparts up to BH masses of $M_{\mathrm{BH, initial}} \approx 10^{6.9}$~M$_{\odot}$, beyond which we are limited by low number statistics. Compared to LTGs, the degree of enhanced BH growth among unstable misaligned ETGs is more modest at ($\sim0.4$~dex) up to BH masses of $M_{\mathrm{BH, initial}} \lesssim 10^{7.3}$~M$_{\odot}$.}
    \label{plot: BH growth scatter ETG LTG}
    \end{center}
\end{figure}

In Figure~\ref{plot: BH growth scatter ETG LTG} we show the growth of the BH mass for samples of ETGs (top) and LTGs (bottom) analogous to Figure~\ref{plot: BH growth scatter}. Galaxies are classified as either ETGs or LTGs using $\kappa_{\rm{co}}^*$ at each snipshot within the $\sim0.5$~Gyr window. We only include galaxies for which this morphological classification remains constant, and thus exclude galaxies that undergo a morphological transformation. For ETGs (LTGs), this returns sub-samples of 2159 (2885) galaxies, of which 285 (64) are experiencing an unstable misalignment and 374 (116) are in a steady-state counter-rotating configuration.

We find the degree of BH growth enhancement varies with galaxy morphology. This enhanced growth is stronger for LTGs ($\sim1.0$~dex) than for ETGs ($\sim0.4$~dex), at least at lower masses where both can be compared. At higher BH masses, our analysis is limited to ETGs given the low number statistics of unstable misaligned LTGs beyond $M_{\mathrm{BH, initial}} \gtrsim 10^{6.9}$~M$_{\odot}$. Consistent with results from Section~\ref{results: enhanced BH growth total}, BH growth among unstable misaligned ETGs is comparable to their aligned and counter-rotating counterparts at higher BH masses.

Over the entire range of BH masses considered, unstable misaligned LTGs show by far the largest median BH growth ($33.7$~per~cent), followed by unstable misaligned ETGs ($10.6$~per~cent), and this result is significant (KS-test statistic $=0.41$, p-value $=1.5\times10^{-8}$). Aligned and counter-rotating galaxies show little dependence on morphology with similar BH growths of $\sim3.5$~per~cent. These results suggest misaligned gas is driven more effectively toward the centre of an LTG. This is expected given that a more disc-like stellar mass distribution will exert stronger torques on misaligned components \citep[][]{Tohline1982, LakeNorman1983}. This results in stronger dissipative forces acting on the gas and shorter relaxation timescales in LTGs compared to ETGs \citep[for details see ][]{Baker2024}. 

Given that unstable misalignments are found to be most effective in growing BHs in LTGs, it is worth considering the degree to which unstable misalignments may dominate the AGN population among discy galaxies at low redshifts. In Appendix~\ref{appendix: AGN at z=0.1} we briefly explore the degree to which AGN are more likely to be found in misaligned systems using a sample of galaxies at $z=0.1$. In summary, we find that the presence of an unstable misalignment increases the likelihood of fuelling AGN activity. This trend is significantly weaker for counter-rotation. Misalignments as a whole are only found in a minority ($\approx15$)~per~cent of LTGs with AGN, whereas they are found over half of ETGs with AGN. Therefore, while unstable misalignments can drive efficient BH growth in LTGs in \textsc{eagle}, they are not the dominant driver of AGN in LTGs at $z=0.1$.

\subsubsection{Gas reservoirs and dynamics}\label{results: gas reservoirs and dynamics}

\begin{figure}
    \begin{center}
    \includegraphics[width=8.5cm]{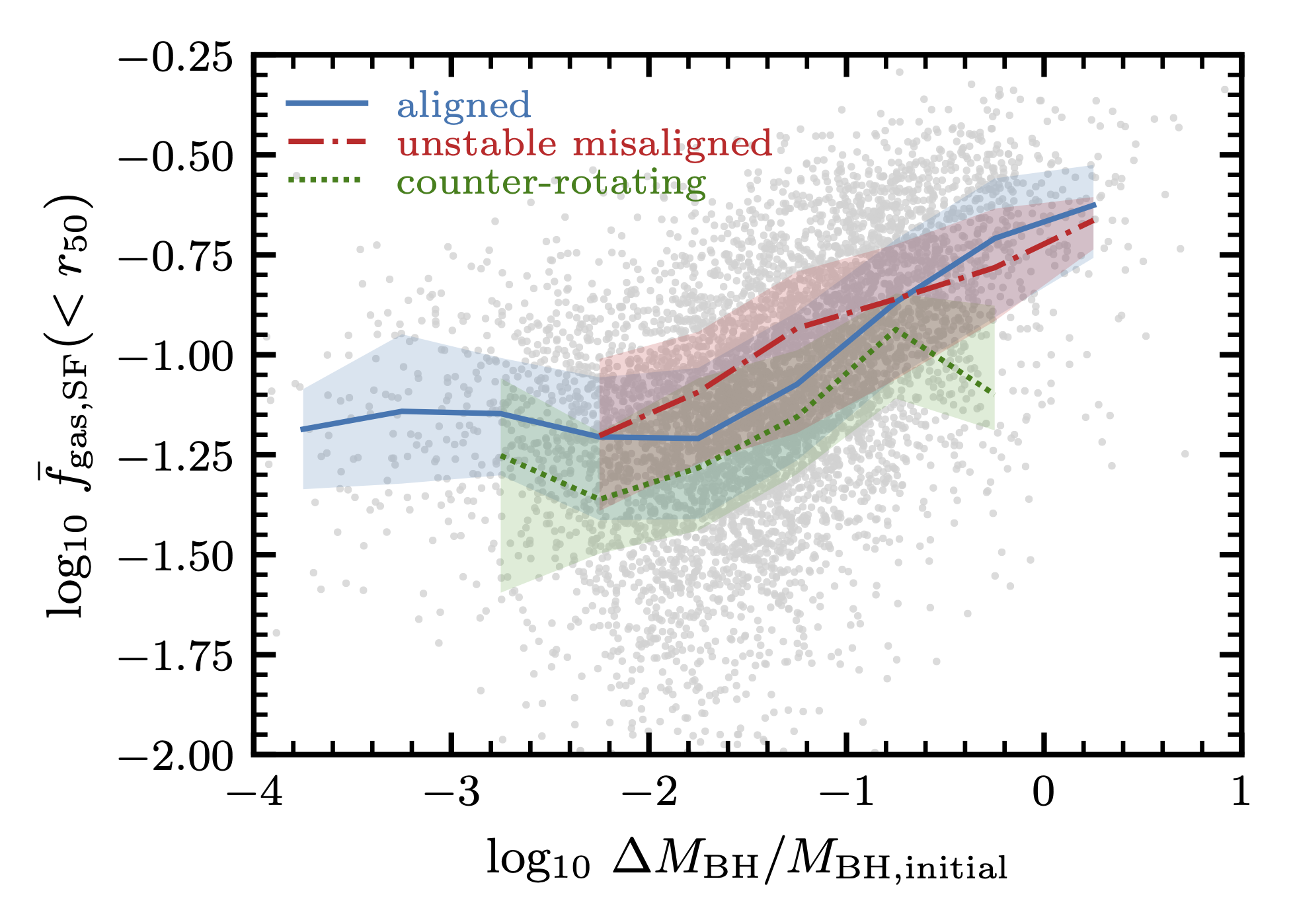}\\
    \caption{Median star-forming gas fraction within $r_{50}$ as a function of fractional BH growth averaged over a $0.50\pm0.05$~Gyr window for sub-samples of aligned (blue solid), counter-rotating (green dotted), and galaxies relaxing from an unstable misalignment (red dash-dotted). Shaded regions show the $25$ and $75$ percentiles. Grey scatter points show the total sample. Galaxies experiencing an unstable misalignment tend to be more gas rich overall, but have similar $f_{\rm{gas,SF}}$ compared to aligned galaxies among the fastest growing BHs.}
    \label{plot: BH growth fgas}
    \end{center}
\end{figure}

In Figure~\ref{plot: BH growth fgas} we show the correlation between average gas$_{\rm{SF}}$ fraction, $f_{\rm{gas,SF}}$, and fractional BH growth over $\sim0.5$~Gyr windows. We find a broadly positive correlation between enhanced BH growth and galaxies that are more gas-rich.

\begin{figure}
    \begin{center}
    \includegraphics[width=8.5cm]{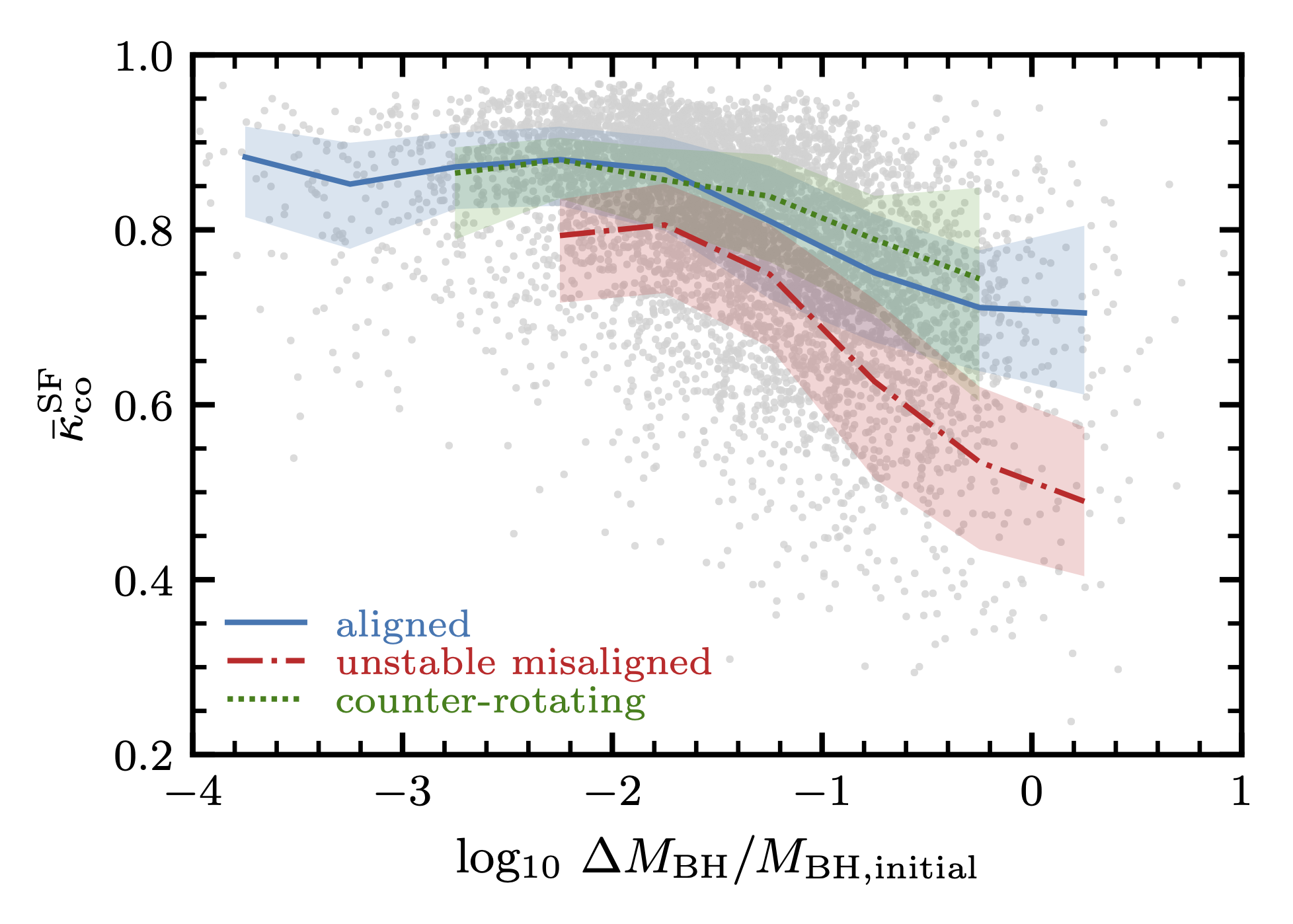}\\
    \caption{Median co-rotational energy fraction for star-forming gas within $r_{50}$ as a function of fractional BH growth averaged over a $0.50\pm0.05$~Gyr window for sub-samples of aligned (blue solid), counter-rotating (green dotted), and galaxies relaxing from an unstable misalignment (red dash-dotted). Shaded regions show the $25$ and $75$ percentiles. Grey scatter points show the total sample. Galaxies experiencing an unstable misalignment tend to have \gassf discs with less rotational support compared to aligned and counter-rotating galaxies.}
    \label{plot: BH growth kappa}
    \end{center}
\end{figure}

On average, galaxies experiencing an unstable misalignment (shown in red) tend to be more gas-rich within $r_{50}$ (median $\bar{f}_{\rm{gas,SF}}=0.123$) compared to aligned galaxies (median $\bar{f}_{\rm{gas,SF}}=0.084$). We note that this may be a selection effect as our sample does not include misalignments with $t_{\rm{relax}}< 0.45$~Gyr which are common among more gas-poor galaxies and which dominate the overall relaxation time distributions \citep[][]{Baker2024}. Nonetheless, given that misalignments can be associated with gas replenishment \citep[e.g.][]{Khim2021, Casanueva2022, Baker2024}, this result is not entirely surprising. 

However, among galaxies with the largest BH growth, unstable misaligned galaxies have comparable gas$_{\rm{SF}}$ fractions. Likewise, among gas rich-galaxies ($\bar{f}_{\rm{gas,SF}}>0.1$), unstable misaligned galaxies continue to show boosted BH growth compared to aligned galaxies (not shown). This hints at an underlying difference in the behaviour of the gas in aligned vs. relaxing galaxies in addition to the degree of gas rich-ness of galaxies.

AGN fueling requires both the presence of gas in the galaxy and a means to transport it to the inner regions of the galaxy (e.g. loss of angular momentum) in the vicinity of the BH \citep[][]{StorchiSchnorr2019}. In Figure~\ref{plot: BH growth kappa} we show the correlation between the co-rotational energy fraction for gas$_{\rm{SF}}$ ($\kappa_{\rm{co}}^{\rm{SF}}$) and fractional BH growth, averaged over $\sim0.5$~Gyr windows. 

We find that BH growth tends to be enhanced in systems with a lower $\kappa_{\rm{co}}^{\rm{SF}}$, regardless of kinematic (mis-)alignment. This may be indicative of greater contributions from non-circular motions (i.e. radial inflows), or the increased presence of \gassf particles in circular orbits that are non-coplanar. Aligned and counter-rotating galaxies show a noticeable decrease ($\Delta \bar{\kappa}_{\rm{co}}^{\rm{SF}}\sim-0.2$) in the average \gassf co-rotational energy fraction over the range of BH growth fractions considered. Among the fastest growing BHs, median $\bar{\kappa}_{\rm{co}}^{\rm{SF}}$ values of $\approx0.70$ indicate that, despite the loss of rotational-support, gas discs in aligned galaxies remain thin and discy. 

As seen in Figure~\ref{plot: BH growth kappa}, we find that the gas discs of galaxies experiencing an unstable misalignment show weaker (co-)rotational support compared to aligned and counter-rotating galaxies with similar BH growth. Consequently, the median \gassf co-rotational energy fraction for unstable misaligned galaxies is lower ($\bar{\kappa}_{\rm{co}}^{\rm{SF}}\approx0.66$) compared to that of aligned ($\bar{\kappa}_{\rm{co}}^{\rm{SF}}\approx0.84$) and counter-rotating galaxies ($\bar{\kappa}_{\rm{co}}^{\rm{SF}}\approx0.82$). Unstable misaligned galaxies also show a larger decrease ($\Delta\bar{\kappa}_{\rm{co}}^{\rm{SF}}\sim-0.3$) in the average co-rotational energy fraction over the range of BH growth within the sample compared to aligned galaxies. Morphologically, this approximately indicates a transition from thin gas discs at low BH growth fractions to thicker gas discs with more dispersion at the highest BH growth fractions \citep[see also][]{Hill2021}.

The weaker (co-)rotational support in gas discs among galaxies experiencing unstable misalignments is unsurprising. During the initial stages of misalignment formation, interactions between any in-situ co-rotating gas and stochastic accretion of misaligned gas naturally increase the dispersion within the gas disc while it settles into a more coherent disc. These interactions have been shown to be an efficient way for gas to dissipate angular momentum \citep[e.g.][]{Sales2012, Taylor2018, Starkenburg2019}. This effect is strongest in the initial snipshots of the $\approx0.5$~Gyr window and likely accounts for many of the systems with relatively low values of $\bar{\kappa}_{\rm{co}}^{\rm{SF}}$. However, once a misaligned gas disc forms and for the remainder of the unstable misalignment, decreased values of $\kappa_{\rm{co}}^{\rm{SF}}$, in excess of aligned galaxy counterparts, are likely indicative of interactions between adjacent rings of non-coplanar gas \citep[see e.g.][]{VoortDavis2015}. Beyond the initial phase of misalignment formation, this is expected to become the dominant method of angular momentum dissipation which allows the gas disc to relax into the galactic plane. The lower values of $\bar{\kappa}_{\rm{co}}^{\rm{SF}}$ in galaxies experiencing an unstable misalignment suggests that misaligned gas is able to efficiently dissipate angular momentum. This causes the central gas density to increase (as evidenced by the relative absence of gas-poor unstable misaligned galaxies in our sample), resulting in higher sub-grid BH accretion rates. 

Although this is one interpretation, an alternative explanation for the weaker values of $\kappa_{\rm{co}}^{\rm{SF}}$ among BHs with larger mass increases may come, at least in part, from the AGN feedback implementation in \textsc{eagle} \citep[for details see][]{SchayeCrain2015}. The stochastic thermal heating of gas particles in the immediate vicinity of a fast-growing BH may increase the velocity dispersion of gas and subsequently decrease the \gassf co-rotational energy fraction. However, this effect is likely limited to the innermost regions of the galaxy. Furthermore, we tend to see $\kappa_{\rm{co}}^{\rm{SF}}$ increase alongside the increased BH growth following misalignment formation as the gas re-forms a disc. This suggests that this effect does not dominate the low values of $\bar{\kappa}_{\rm{co}}^{\rm{SF}}$ seen in galaxies experiencing an unstable misalignment. We discuss this further in Section~\ref{discussion:}.

In contrast to the unstable misaligned sample, continuously counter-rotating galaxies tend to contain gas discs that are dynamically relaxed into thin discs ($\bar{\kappa}_{\rm{co}}^{\rm{SF}}\gtrsim0.7$). In the absence of external factors, gas particles in retrograde orbits must therefore rely on sub-grid stellar wind modelling to induce change in the angular momentum. Stellar-gas counter-rotation in \textsc{eagle} tends to be long-lived, forming a significant fraction of the galaxy population by $z=0.1$ \citep[for details, see][]{Baker2024}. While the fraction of counter-rotating galaxies is comparable to observations, the stellar mass loss model in \textsc{eagle} is unable to accurately capture the long-term angular momentum dissipation of a counter-rotating gas disc imbedded in a co-rotating stellar disc. This is likely a result of the limited resolution of large-scale cosmological simulations which results in similar growth of BHs in aligned and counter-rotating galaxies (see Figure~\ref{plot: BH growth scatter}).

\subsection{BH growth history}\label{results: BH growth history at z=0.1}

\subsubsection{Overmassive BHs at z = 0.1}\label{results: overmassive BHs at z=0.1}

\begin{figure}
    \begin{center}
    \includegraphics[width=8.5cm]{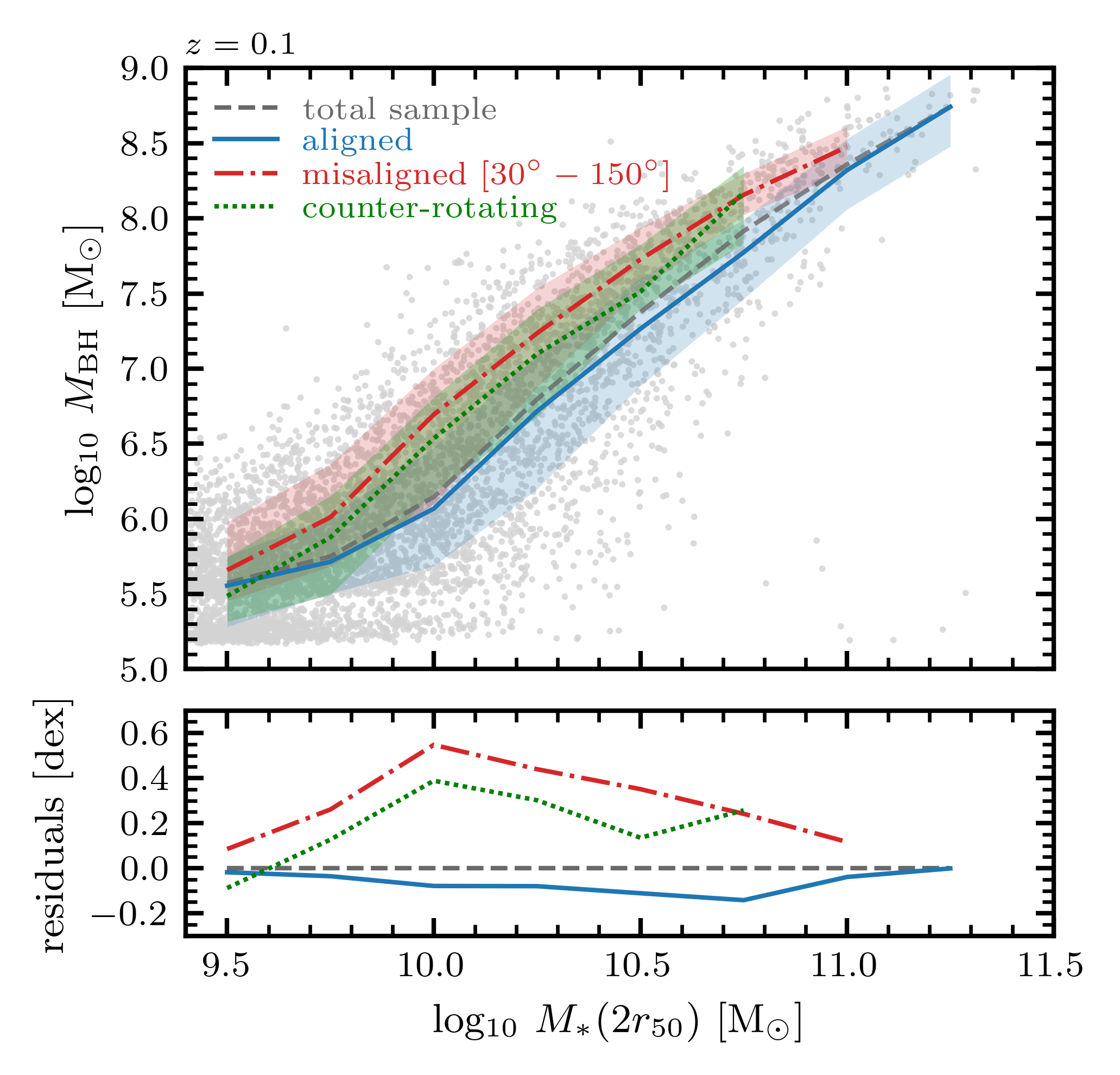}\\
    \caption{$M_* - M_{\mathrm{BH}}$ relation at $z=0.1$ for galaxies in various stellar-gas kinematic states. Top: median lines for the total sample (black dashed) and sub-samples of aligned (blue solid), unstable misaligned (red dash-dotted), and counter-rotating (green dotted) galaxies for bins with a minimum bin count of 10. Shaded regions show the $25$ and $75$ percentiles for aligned, unstable misaligned, and counter-rotating galaxies. Grey scatter points show the total sample. Bottom: residuals of aligned, unstable misaligned, and counter-rotating medians, with respect to the total sample. We find overmassive BHs preferentially reside in galaxies that are currently in the unstable misaligned or counter-rotating regime, with a maximum BH mass difference of $\sim0.6$~dex at a turnover mass of $\sim10^{10}$~M$_{\odot}$.}
    \label{plot: BH overmassive z=0.1}
    \end{center}
\end{figure}

It is worth investigating whether the enhanced BH growth found in galaxies experiencing unstable misalignments is imprinted on the galaxy population at $z=0.1$. This redshift was chosen arbitrarily and lies within the redshift limit probed by recent IFU surveys such as SAMI and Mapping Nearby Galaxies at Apache Point Observatory \citep[MaNGA; ][]{Bundy15}. Given the relatively low numbers of misaligned galaxies that coincide with this redshift in our pre-existing sample \citep[due to the relaxation requirement, see][]{Baker2024}, we extract a new sample of 5372 galaxies at $z=0.1$ matching the criteria in Section~\ref{Sample:} with no minimum BH mass requirement. These galaxies are classified according to their current kinematic state at $z=0.1$, of which 4166 are aligned ($\psi_{\rm{3D}}<30^{\circ}$), 308 are counter-rotating ($\psi_{\rm{3D}}>150^{\circ}$), and 898 are in the unstable misaligned regime ($30^{\circ}<\psi_{\rm{3D}}<150^{\circ}$). 

In Figure~\ref{plot: BH overmassive z=0.1} (top) we show the $M_* - M_{\mathrm{BH}}$ relationship of this sample alongside medians for each sub-sample of kinematic state. Stellar masses are measured within $2r_{50}$. Residuals of these medians are shown with respect to the total sample (bottom). We find overmassive black holes preferentially residing in galaxies that are currently misaligned and, to a lesser degree, counter-rotating. This trend is seen most strongly for intermediate-mass galaxies with $10^{9.7} \lesssim M_*/\rm{M}_{\odot} \lesssim 10^{10.5}$. For instance, misaligned and aligned galaxies show a maximum difference of $\sim0.6$~dex between BH masses at a turnover mass of $M_*\sim10^{10}$~M$_{\odot}$. Comparing the distributions of $M_{\rm{BH}}/M_*$ for aligned and misaligned galaxies, we find the differences to be significant (KS-test statistic $=0.21$, p-value $\ll 10^{-10}$). Similarly, for aligned and counter-rotating galaxies (KS-test statistic $=0.27$, p-value $\ll 10^{-10}$). As before, this enhancement is seen more strongly among LTGs than in ETGs. These results remain unchanged if we crudely mimic observational measurements by using projected angles in two-dimensional (2D) space instead of 3D angles.

\begin{figure}
    \hspace*{0cm}
    \includegraphics[width=8.5cm]{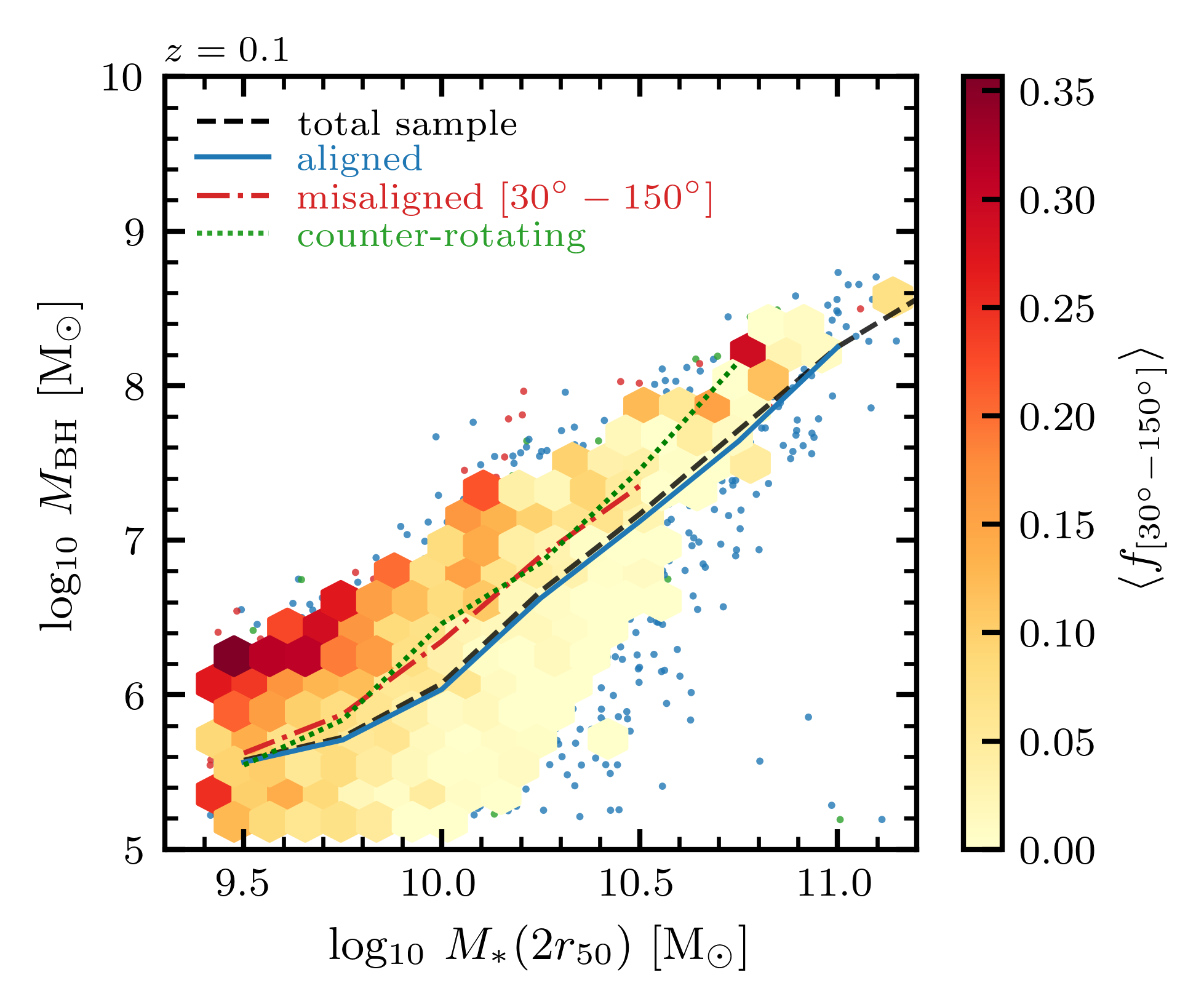}
    \caption{$M_* - M_{\mathrm{BH}}$ relation at $z=0.1$ for a representative sample of galaxies with reliable kinematic measurements over $0.10\lesssim z\lesssim0.28$. Bins coloured according to the fractional time a galaxy spends in the unstable misaligned regime ($f_{\mathrm{[30^{\circ}-150^{\circ}]}}$) within this redshift range, averaged for all galaxies within a given bin (minimum count of 5). Medians are shown for the total sample (black dashed) and sub-samples of aligned (blue solid), unstable misaligned (red dash-dotted), and counter-rotating (green dotted) galaxies for bins with a minimum bin count of 10. We find a strong correlation between galaxies hosting overmassive BHs and galaxies with a history of spending more time in the unstable misaligned regime.}
    \label{plot: BH overmassive hexbin}
\end{figure}

Overmassive BHs are also found in counter-rotating systems at $z=0.1$, but with a weaker deviation from the total population. Counter-rotating galaxies are commonly formed from the relaxation of past unstable misalignments \citep[e.g.][]{Baker2024}, assuming the counter-rotating gas disc did not form directly from retrograde gas accretion onto a gas-poor galaxy. As such, many of these present-day counter-rotating systems may have experienced a past unstable misalignment and corresponding phase of enhanced BH accretion. While their current BH growth may be indistinguishable from the aligned galaxy population, counter-rotating galaxies in \textsc{eagle} encapsulate their past misalignment history in the form of overmassive BHs.

\subsubsection{Galaxy kinematic history at z = 0.1}\label{results: Galaxy kinematic history at z=0.1}

We can also investigate the degree to which the mass of the BH traces the history of past unstable misalignments. Beginning with our sample of 5372 galaxies at $z=0.1$, we select a subset of 3931 galaxies that follow the selection criteria of \citet{Baker2024} for reliable kinematic measurements over the preceding $2$~Gyr ($0.10\lesssim z\lesssim0.28$). Of this subset, 3367 are aligned, 200 are counter-rotating, and 364 are in the unstable misaligned regime. With the exception of low-mass galaxies with $M_*\lesssim10^{9.7}$~M$_{\odot}$, this creates a sample of galaxies with stellar masses representative of those at $z=0.1$.

In Figure~\ref{plot: BH overmassive hexbin} we show the relationship between galaxies with overmassive BHs and their stellar-gas kinematic histories over the past $2$~Gyr. We denote the fraction of time a galaxy has spent in the unstable misaligned regime over the window $0.10\lesssim z\lesssim0.28$ as $f_{\mathrm{[30^{\circ}-150^{\circ}]}}$. We find a clear trend between galaxies hosting overmassive BHs and the average time spent in the unstable misaligned regime. This fraction tends to increase with distance offset from the median $M_*-M_{\mathrm{BH}}$ relationship. Correspondingly, galaxies with undermassive BHs rarely show any sign of past unstable misalignments within $0.10\lesssim z\lesssim0.28$. These trends are seen more strongly in galaxies with $M_*\lesssim10^{10.3}$~M$_{\odot}$.

We also find a trend between overmassive BHs and a rich history of past counter-rotation (not shown), although this trend is weaker compared to that of unstable misalignments. Similarly for the misaligned result in Figure~\ref{plot: BH overmassive hexbin}, this is seen most strongly for intermediate-mass galaxies with $M_*\sim10^{10}$~M$_{\odot}$. As explained above, this result is unsurprising given stellar-gas counter-rotation can be the result of a past relaxation. We find that both trends are also seen for galaxy histories extending over the past $4$~Gyr ($0.10\lesssim z\lesssim0.51$) and $6$~Gyr ($0.10\lesssim z\lesssim0.85$), although these are not shown. These samples tend to exclude an increasing number of low-mass galaxies as lookback time increases, and thus we caution that these longer timescale results are less certain due to lower number statistics. 

Furthermore, by only selecting the subset of galaxies that are aligned at $z=0.1$, we continue to find that overmassive BHs trace galaxies with a history of more unstable misalignments (not shown). Indeed, we find the past kinematic (mis)alignment history of a galaxy is a significantly stronger indicator of the relative mass of present-day BHs than compared to gas-richness ($f_{\rm{gas,SF}}>0.1$) history. Therefore, at least in \textsc{eagle}, the relative deviation of a BH from the $M_*-M_{\mathrm{BH}}$ relationship at $z=0.1$ can provide clues about the past kinematic history of gas within the galaxy, regardless of the current state of stellar-gas (mis)alignment.

Observational studies of local galaxies have shown that overmassive BHs tend to preferentially reside in ETGs \citep[e.g.][]{GrahamSahu23}. This has been attributed to the cumulative effects of mergers acting as a source of both cold gas to fuel BH growth and disruption of pre-existing cold gas reservoirs during the assembly of these galaxies. Our findings suggest that the presence and persistence of an unstable kinematic misalignment is a complementary channel for galaxies to acquire an overmassive BH without the prerequisite of a galaxy merger. Furthermore, the presence of an undermassive BH is a strong indicator of past stellar-gas alignment. Given both mergers and stellar-gas kinematic misalignments preferentially occur among ETGs compared to LTGs in \textsc{eagle} \citep[e.g.][]{Casanueva2022, Baker2024}, this further complements the observed relationship between galaxy morphology and location on the $M_*-M_{\mathrm{BH}}$ plane. We note that observational measurements of single-epoch black hole masses using the broad H$\beta$ line typically have mass uncertainties of $0.3-0.4$~dex \citep[e.g.][]{DallaBonta20}. As such, the difference of $\sim0.6$~dex we find between intermediate-mass misaligned and aligned galaxies is likely to be detected in observations.

\section{Discussion}\label{discussion:}

Overall, our results are consistent with the picture that the  persistence of an unstable stellar-gas kinematic misalignment is an effective mechanism to drive gas inwards for increased BH activity. This is in good agreement with results from other cosmological simulations \citep[e.g.][]{Starkenburg2019, Khoperskov2020, Duckworth2020a, Duckworth2020b} and recent observational findings \citep[e.g.][]{Ruffa2019b, Raimundo2017, Raimundo2021, Raimundo2023}. 

In previous work, \citet{Duckworth2020b} used the IllustrisTNG-100 simulation and a sample of $\sim2500$ galaxies (with masses of $10^{8.5}<M_*/\mathrm{M}_{\odot}<10^{11.5}$) to investigate AGN activity in misaligned galaxies. This was done through the construction of mock observational images to mimic results from MaNGA, with 2D position angles (PAs) used to classify galaxies as misaligned ($\Delta\mathrm{PA}\geq30^{\circ}$). In agreement with our results, low-mass ($M_*<10^{10.2}$~M$_{\odot}$) misaligned galaxies show boosted AGN luminosity and BH growth at $z\sim0.1$ with results becoming less conclusive in higher-mass galaxies. Similarly, \citet{Khoperskov2020} find that non-aligned gas is efficiently funneled inward in IllustrisTNG through interactions with aligned gas. We note that IllustrisTNG uses a 2-mode AGN feedback model, with $M_{\rm{BH}}\sim10^{8}$~M$_{\odot}$ corresponding to the typical transition between thermal and a more violent kinetic-mode feedback. The implementation of such a feedback mode can have a strong effect on the gas content of galaxies, especially at the more massive end \citep[e.g.][]{Dave2020}. As such, while our results may be sensitive to the BH feedback model used, it is reassuring that we find good qualitative agreement between the two simulations within the range of galaxy masses considered for this paper. A detailed comparison between massive galaxies (with correspondingly massive BHs), in which the feedback regime may have a stronger impact on our results, is strongly limited by our small sample of $M_{\rm{BH}}\sim10^{8}$~M$_{\odot}$ BHs in misaligned galaxies.

Counter-rotating systems in IllustrisTNG have also been found with boosted AGN luminosities and BH growth \citep[][]{Khoperskov2020, Duckworth2020b}. While we do find overmassive BHs in counter-rotating galaxies at $z=0.1$ in \textsc{eagle}, we do not find significantly enhanced BH growth or luminosity in these systems. Instead, these overmassive BHs trace prior unstable misalignments. This discrepancy may be due to differences in the sub-grid models and underlying hydrodynamic scheme of these simulations. However, our results are supported by recent observational results that found no evidence of enhanced AGN luminosities among AGN host galaxies with larger misalignment angles \citep[][]{Winiarska2025}. 

Misaligned gas has also been associated with lower angular momentum in simulations \citep[e.g.][]{VoortDavis2015, Starkenburg2019, Duckworth2020a} and recent observational studies \citep[e.g.][]{Xu2022}. Specifically for \textsc{eagle}, \citet{Casanueva2022} find misaligned galaxies have star-forming discs that are more compact than aligned counterparts. This hints at the efficiency of misaligned gas to lose angular momentum as it relaxes into a dynamically-stable configuration, driving gas inward. Again, our results are in good qualitative agreement with this interpretation.

Galaxy mergers have long been established as an effective means for gas to dissipate angular momentum and potentially fuel AGN activity \citep[see][ and references therein]{StorchiSchnorr2019}. For instance, mergers have been found as a key driver of overmassive BHs in \textsc{eagle} through the disruption of pre-existing co-rotating gas \citep[e.g.][]{Davies2022, Davies2024}. Likewise, mergers have been found to increase the incidence of luminous AGN in \textsc{eagle} \citep{McAlpine2020}. As such, it is important to consider any bias associated with our sample of relaxing unstable misalignments as these may be associated with significantly more mergers. 

For this we compare the incidence of mergers with stellar mass-ratios of $>1/10$ within our $\sim0.5$~Gyr windows between our samples of aligned, unstable misaligned, and counter-rotating galaxies. We find a higher incidence of mergers in our unstable misaligned sample ($\approx9$~per~cent) compared to our aligned ($\approx4$~per~cent) and counter-rotating samples ($\approx3$~per~cent). This is consistent with previous results in cosmological simulations that found mergers to be sub-dominant in the formation of misalignments \citep[e.g.][]{Khim2021, Baker2024}. Our results remain unchanged if we remove \textit{all} unstable misaligned galaxies that experience a merger within our $\sim0.5$~Gyr windows from our sample. As BH growth has been shown to increase significantly within $\pm0.12$~Gyr of coalescence \citep[see][]{McAlpine2020}, we reason that the lack of change in our results emphasises the dominance of misalignment-driven growth over that of in-plane gas inflow driven by galaxy mergers in our sample. We note that this does not preclude a higher incidence of galaxy interactions within the unstable misaligned sample. While a higher incidence of interactions have been shown to increase BH luminosities, the increase is significantly weaker than for mergers \citep{McAlpine2020}. Therefore, we reason that a $\sim5$~per~cent increase in the incidence of mergers between aligned and relaxing unstable misaligned galaxies is unable to account for the systematically higher BH growth found in Figure~\ref{plot: BH growth scatter}. We conclude that the boosted BH growth seen in our results is likely not due to a higher incidence of mergers in misaligned galaxies \citep[see also][]{Raimundo2025}. 

As explored in \citet{McAlpine2020}, despite galaxy mergers increasing the incidence of luminous AGN, merger-driven BH growth was found to contribute only $\sim15$~per~cent toward present-day BH mass in \textsc{eagle}. Similar results have also been found in the \textsc{horizon-agn} \citep[][]{Dubois14} simulation \citep[see][]{Martin2018}. Therefore, it is worth considering how misalignment-driven growth may compare. Compared to a control sample, \citet{McAlpine2020} report a $2-3$ times increase in BH accretion rates in merging systems. For comparison, we report an increase of $\sim4$ times ($\sim0.6$~dex) among unstable misaligned galaxies compared to aligned counterparts. In both cases, BH growth was found to be increased predominantly in less massive BHs ($M_{\rm{BH}}\lesssim10^{7}$~M$_{\odot}$). Unstable misalignments in \textsc{eagle} were found with median lifetimes of $\sim0.5$~Gyr, but with a large standard deviation of $\sim0.6$~Gyr \citep[see][]{Baker2024}. Furthermore, misalignments have been found with diverse formation pathways in which mergers are sub-dominant \citep[e.g.][]{Khim2021, Casanueva2022, Baker2024}. As such, misalignments not only provide more BH growth than mergers with the ability to spend considerable ($\gtrsim0.5$~Gyr) time with these features, but can also occur more frequently and independently of galaxy mergers. While quantifying in detail the contribution of misalignments toward present-day BH mass is beyond the scope of this paper, we suggest that misalignments may contribute more toward present-day BH mass in \textsc{eagle} than mergers alone.

AGN feedback is likely to play a role in reducing the degree of BH growth in more massive misaligned systems. As summarised in \citet{Mcalpine2018}, lower-mass BHs ($M_{\mathrm{BH}}\lesssim10^{7}$~M$_{\odot}$) tend to grow rapidly as AGN feedback is unable to self-regulate gas inflows, leading to a rise in central gas density \citep[see also][]{Bower2017}. In more massive BHs ($M_{\mathrm{BH}}\gtrsim10^{7}$~M$_{\odot}$), AGN feedback is able to self-regulate inflows and growth. As such, we speculate that the increased efficiency of AGN feedback for higher-mass BHs is likely reducing the efficiency of unstable misalignments to enhance BH growth over the time periods considered in this work. Consequently, this is imprinted on the $M_* - M_{\mathrm{BH}}$ relationship at $z=0.1$ which results in a weaker BH mass offset between misaligned and aligned galaxies for higher-mass galaxies (see Figure~\ref{plot: BH overmassive z=0.1}).

Finally, while the increased AGN feedback in misaligned systems may be contributing to the lower gas co-rotational energy fraction, we do not believe it is the key driver of the unstable misalignments over $\sim0.5$~Gyr periods. This is because the angular momentum of the outflows would need to dominate over the angular momentum of the remaining gas disc. This would imply that a large amount of \gassf would be carried away within $r_{50}$, starving the BH of long-term fuel. Yet, we continue to clearly identify enhanced BH growth in misaligned galaxies for the largest time window considered ($\approx1.0$~Gyr). Secondly, the thermal BH feedback implementation in \textsc{eagle} means outflows naturally follow the path of least resistance perpendicular to the gas disc \citep{Hartwig18, Mitchell2020outflow}. Thus, it may be difficult for AGN-driven outflows to misalign the gas disc. This is supported by the finding that wind recycling rates in \textsc{eagle} tend to be low and that outflows at galactic scales tend to be ejective out to high radii \citep[][]{Mitchell2020outflow, Mitchell2020inflow}. In contrast, outflows in IllustrisTNG galaxies are more violent and tend to follow ballistic orbits \citep[][]{Mitchell2020outflow} which manifest as galactic fountains \citep[][]{ShapiroField76}. While this makes AGN-driven misalignments more feasible in IllustrisTNG \citep[see also][]{Duckworth2020b}, we argue that this process is likely rare in \textsc{eagle}. However, we do not rule out the possibility that significant AGN feedback preceded the misalignment, decreasing the gas fraction and easing the formation of a misalignment \citep[as proposed by e.g.][]{Casanueva2022, Cenci2023}. Or, alternatively, that ongoing AGN feedback is disrupting the axis of gas accretion onto the misaligned gas disc as suggested by previous authors \citep[e.g.][]{Starkenburg2019, Khim2021, Casanueva2022}.


\section{Conclusions}\label{conclusion:} 

In this work, we have used the \textsc{eagle} cosmological simulations to investigate the growth of the central BH in galaxies with stellar-gas kinematic misalignments between $0<z<1$. We used a sample of 5570 galaxies with masses of $>10^{9.5}$~M$_{\odot}$ and reliable kinematic and BH mass measurements. We traced the change in BH mass over $\approx0.5$~Gyr windows in galaxies experiencing an unstable kinematic misalignment ($30^{\circ} \leq \psi_{\mathrm{3D}} \leq 150^{\circ}$) and compared these to a sample of steady-state aligned ($\psi_{\mathrm{3D}}<30^{\circ}$) and counter-rotating ($\psi_{\mathrm{3D}}>150^{\circ}$) galaxies. Using this sample, we established correlations between the amount and kinematics of star-forming gas and the degree of BH growth. Finally, we used a sample of 3931 galaxies at $z=0.1$ with reliable kinematic measurements over the previous $2$~Gyr to investigate the relationship between overmassive BHs and the stellar-gas kinematic history of galaxies. Our results can be summarised as follows:

\begin{enumerate}
\item BHs residing in galaxies experiencing an unstable misalignment experience significantly enhanced ($\sim0.6$~dex) growth and accretion rates, growing their BH masses by $\approx13$~per~cent over $\approx0.5$~Gyr windows. In comparison, aligned ($<30^\circ$) and counter-rotating ($>150^\circ$) galaxies grow their BHs by $\approx3.9$~per~cent and $\approx3.4$~per~cent over the same time period, respectively. This boosted BH growth ($\sim0.6$~dex) is seen most strongly for BHs of mass $\lesssim10^{7.2}$~M$_{\odot}$, weakening at higher BH masses (see Figures~\ref{plot: BH growth scatter}~and~\ref{plot: BH growth hist}).

\item The degree of enhanced BH growth depends on morphology only in galaxies experiencing an unstable misalignment. Unstable misaligned LTGs provide a \textit{greater} enhancement ($\sim1.0$~dex) to their BH growth than unstable misaligned ETGs ($\sim0.4$~dex) over $\approx0.5$~Gyr windows. Consequently, unstable misaligned LTGs show the greatest BH mass increase ($\approx34$~per~cent), followed by unstable misaligned ETGs ($\approx11$~per~cent), with the remaining aligned/counter-rotating LTGs and ETGs experiencing typical BH mass increases of $\sim3.5$~per~cent over $\approx0.5$~Gyr windows. We attribute this to the greater torques experienced by the misaligned gas from more discy stellar distributions, resulting in stronger angular momentum dissipation and inflow of gas (see Figure~\ref{plot: BH growth scatter ETG LTG}). 

\item While our sample of unstable misaligned galaxies tend to be more gas-rich than aligned or counter-rotating galaxies, we find only a weak correlation between BH growth and star-forming gas fraction irrespective of kinematic (mis)alignment (see Figure~\ref{plot: BH growth fgas}).

\item Star-forming gas in galaxies experiencing an unstable misalignment has less rotational-support. This is likely indicative of the angular momentum dissipation experienced by misaligned gas, driving gas inward to be accreted by the central BH as the disc relaxes into the galactic plane (see Figure~\ref{plot: BH growth kappa}).

\item At $z=0.1$, BH masses are larger ($\sim0.6$~dex) among intermediate-mass galaxies with unstable misalignments compared to aligned counterparts (see Figure~\ref{plot: BH overmassive z=0.1}). Likewise, the population of overmassive BHs at $z=0.1$ is dominated by galaxies that have spent more time in the unstable misaligned regime over the past $2$~Gyr ($0.10\lesssim z\lesssim0.28$, see Figure~\ref{plot: BH overmassive hexbin}).

\item At $z=0.1$, counter-rotating galaxies tend to host overmassive BHs. This is despite finding similar BH growth between aligned and counter-rotating galaxies. We suggest that the BHs of counter-rotating galaxies were affected by their past unstable misalignment and subsequent enhanced growth, followed by a relaxation into the counter-rotating regime (see Figure~\ref{plot: BH overmassive z=0.1}).

\item At $z=0.1$, galaxies with AGN show a significantly higher fraction ($\sim38$~per~cent) of misaligned ($\psi_{\mathrm{3D}}>30^{\circ}$) galaxies than the total galaxy population ($\sim23$~per~cent). Among galaxies with AGN, unstable misalignments ($30^{\circ} \leq \psi_{\mathrm{3D}} \leq 150^{\circ}$) are found in about half of ETGs but in only $\sim9$~per~cent of LTGs. Despite the efficient BH growth experienced by LTGs with unstable misalignments, we suggest that the relative rarity of misalignments in LTGs results in other processes dominating AGN activity in these systems (see Appendix~\ref{appendix: AGN at z=0.1}).
\end{enumerate}

We conclude that, at least in \textsc{eagle}, the presence of an unstable stellar-gas kinematic misalignment is an effective means to drive gas inward to boost the growth of the central BH and trigger AGN activity. Thus, at least statistically, the relative deviation from the $M_* - M_{\rm{BH}}$ relation could be used to infer a galaxy's past (mis)alignment history.

It is clear that more work is needed to fully understand the connection between AGN activity and the presence of stellar-gas kinematic misalignments. For instance, the degree to which misalignments contribute toward non-merger BH growth over a galaxy's lifetime is an exciting question for future studies. Future observational studies with e.g. the Atacama Large Millimeter/submillimeter Array (ALMA), Multi Unit Spectroscopic Explorer (MUSE) and James Webb Space Telescope (JWST) alongside simulations that resolve the multiphase ISM will allow for more thorough investigations of this subject.


\section*{Acknowledgements}

We thank the anonymous referee for constructive comments and recommendations. We acknowledge support from the UK Science Technologies and Facilities Council (STFC) for PhD studentship funding and support for TAD through grant ST/W000830/1. We acknowledge the Virgo Consortium for making their simulation data available and thank Robert Crain and Liverpool John Moores University for providing the \textsc{eagle} snipshots. The \textsc{eagle} simulations were performed using the DiRAC-2 facility at Durham, managed by the ICC, and the PRACE facility Curie based in France at TGCC, CEA, Bruy\`eres-le-Ch\^atel. FvdV is supported by a Royal Society University Research Fellowship (URF\textbackslash R1\textbackslash 191703 and URF\textbackslash R\textbackslash241005). S. I. R acknowledges support from STFC/UKRI via grant reference ST/Y002644/1. 

\section*{Data Availability}

The \textsc{eagle} simulations and database are publicly available at: http://icc.dur.ac.uk/Eagle/database.php. Detailed guides to simulation parameters, models, and how to access and query data can be found in \citet{EAGLEparticleRef} and \citet{Mcalpine2016}. \textsc{eagle} "snipshots" are available on request from members of the \textsc{eagle} team. 



\bibliographystyle{mnras}
\bibliography{library_paper2} 




\appendix

\section{AGN among misaligned galaxies at z=0.1}\label{appendix: AGN at z=0.1}

As shown in Sections~\ref{results: enhanced BH growth total}~and~\ref{results: enhanced BH growth ETGs and LTGs}, galaxies with unstable stellar-gas kinematic misalignments ($30^{\circ}<\psi_{\mathrm{3D}}<150^{\circ}$) show enhanced BH growth. It is therefore interesting to consider whether we see more AGN signatures at $z=0.1$ in galaxies that are misaligned.

We consider two approaches; firstly, the degree to which AGN fractions are increased among aligned ($\psi_{\mathrm{3D}}<30^{\circ}$), misaligned ($\psi_{\mathrm{3D}}>30^{\circ}$), and counter-rotating galaxies ($\psi_{\mathrm{3D}}>150^{\circ}$); and secondly, the degree to which misalignment and counter-rotating fractions vary among the AGN and total galaxy population. In both cases, we use the pre-existing sample of 5372 galaxies at $z=0.1$ as outlined in Section~\ref{results: overmassive BHs at z=0.1}. We define AGN host galaxies using two separate criteria as commonly used in other studies \citep[e.g.][]{McAlpine2020}. Firstly, using a threshold of $L_{\rm{bol}}>10^{43}$~erg~s$^{-1}$, and secondly, using a threshold of $\lambda_{\mathrm{Edd}}>0.01$. In total, we find 268 $L_{\rm{bol}}$-defined AGN (of which 145 are ETGs and 123 are LTGs), and 299 $\lambda_{\mathrm{Edd}}$-defined AGN (of which 152 are ETGs and 147 are LTGs).

\begin{figure}
    \hspace*{0cm}
    \includegraphics[width=8.5cm]{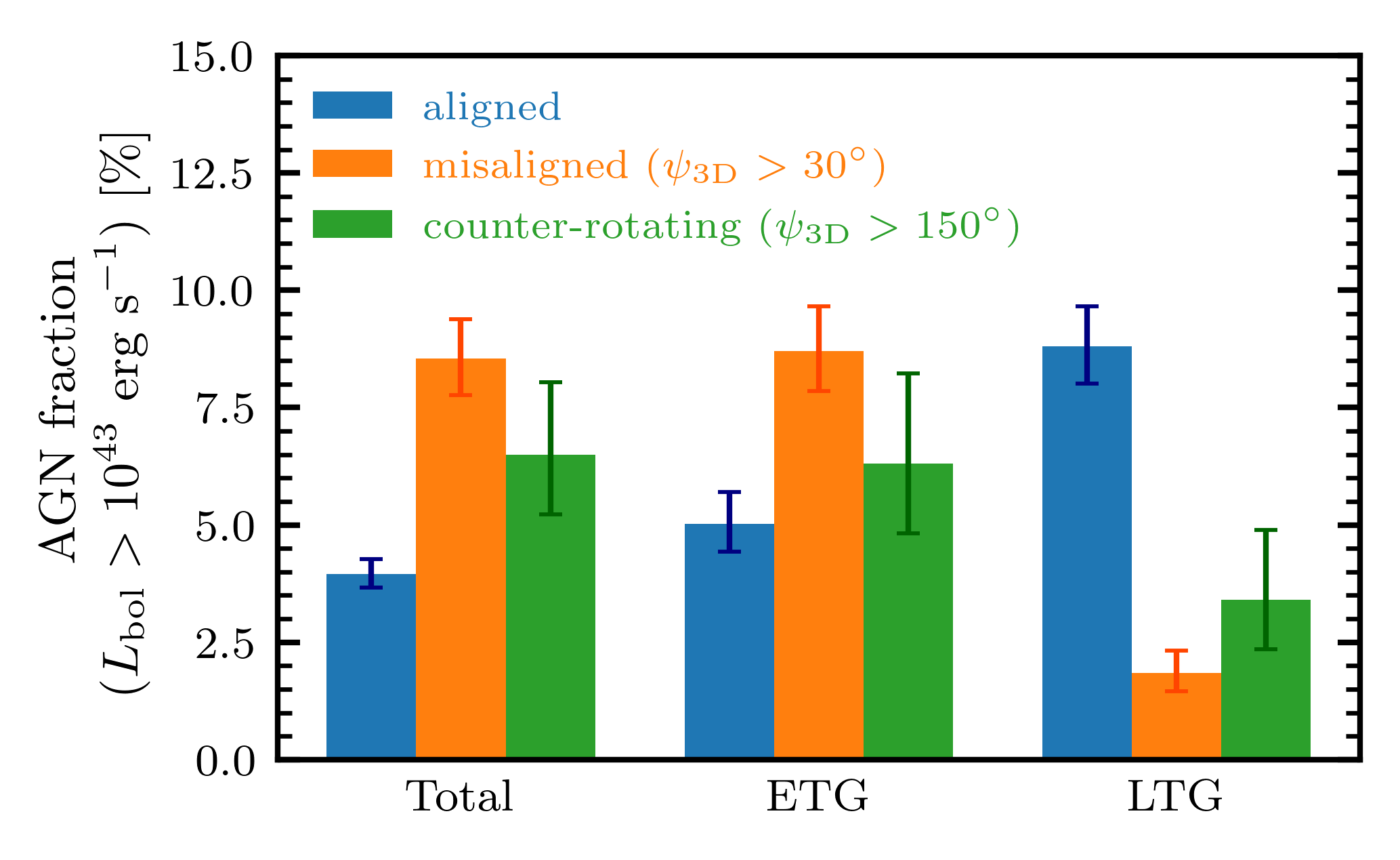}
    \caption{Fraction of AGN (defined as BH luminosities of $L_{\rm{bol}}>10^{43}$~erg~s$^{-1}$) among the total galaxy sample at $z=0.1$ among populations of aligned (blue, first bar of each group), misaligned (orange, second bar of each group), and counter-rotating galaxies (green, third bar of each group). Additionally, we show the fraction of AGN among sub-samples of aligned/misaligned/counter-rotating ETGs and LTGs. The error bars indicate the $1\sigma$ ($68$~per~cent) binomial confidence intervals estimated using the Jeffreys prior. We find higher AGN fractions among misaligned galaxies for the total galaxy population and for ETGs. Conversely, AGN fractions are lowest among misaligned LTGs.}
    \label{plot: f AGN z=0.1}
\end{figure}

In Figure~\ref{plot: f AGN z=0.1} we show the $L_{\rm{bol}}$-defined AGN fraction among galaxies in three states of (mis)alignment for the total sample, and split among ETGs and LTGs. Among all galaxies studied, we see a clear increase in the AGN fraction ($8.5\pm0.8$~per~cent) among misaligned galaxies, compared to aligned galaxies ($4.0\pm0.3$~per~cent). This is also seen for ETGs, with AGN fractions of $8.7\pm1.0$~per~cent and $5.0_{-0.6}^{+0.7}$~per~cent for misaligned and aligned ETGs, respectively. This trend is reversed for LTGs, with higher AGN fractions among aligned ($8.8_{-0.8}^{+0.9}$~per~cent) LTGs than in misaligned LTGs ($1.8_{-0.4}^{+0.5}$~per~cent). These results are all significant at the $2\sigma$ confidence level. Counter-rotating galaxies show AGN fractions intermediate between aligned and misaligned galaxies, but this is only significant to the $1\sigma$ confidence level. Among counter-rotating ETGs and LTGs, we are limited by low number statistics and the increased AGN fraction is not significant. These results are largely unchanged for $\lambda_{\mathrm{Edd}}$-defined AGN (not shown). We note that these states are classified using the galaxy's inherent (i.e. 3D) misalignment angles but that similar results are obtained using projected (i.e. 2D) misalignment angles.

\begin{figure}
    \hspace*{0cm}
    \includegraphics[width=8.5cm]{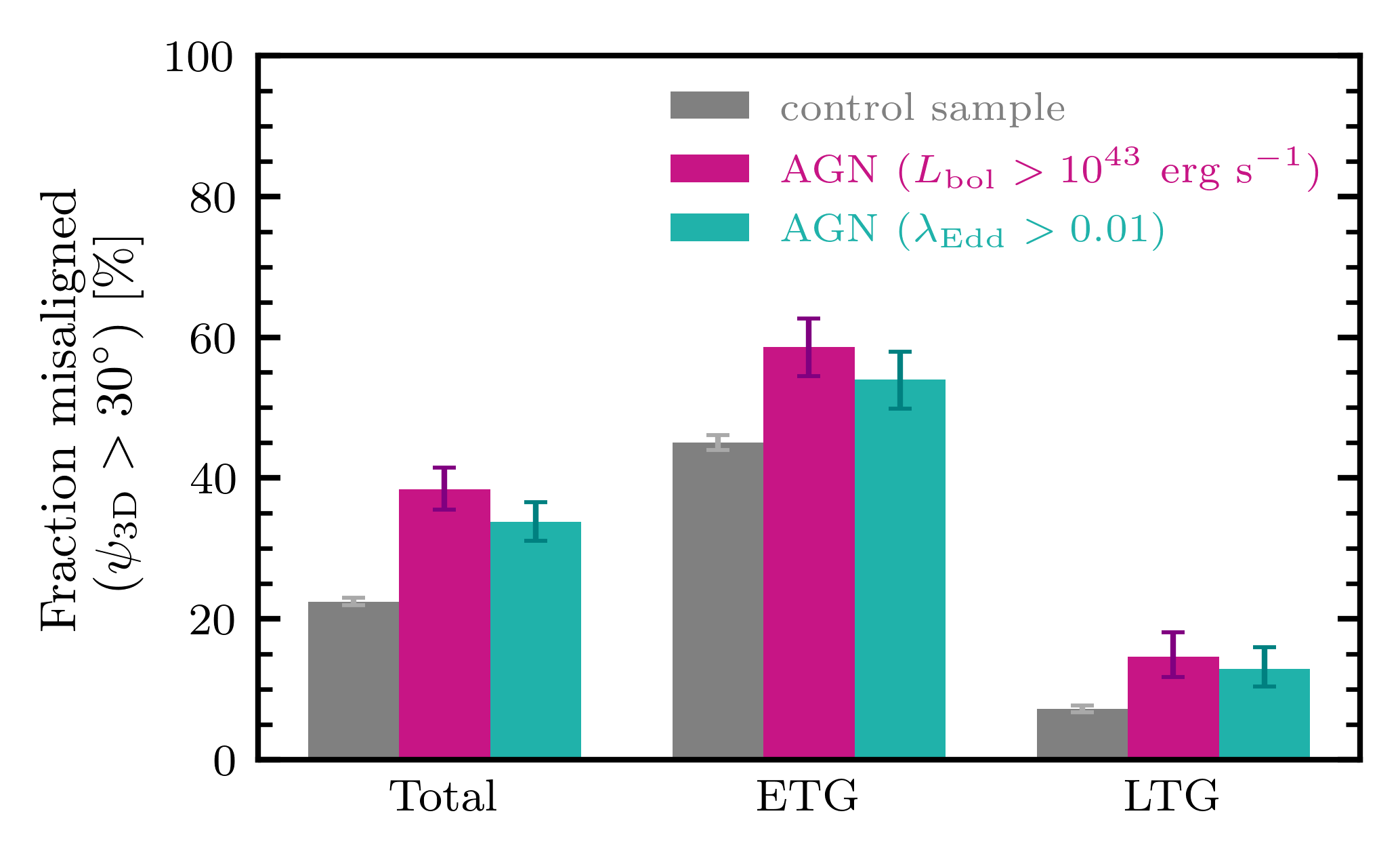}
    \hspace*{0cm}
    \includegraphics[width=8.5cm]{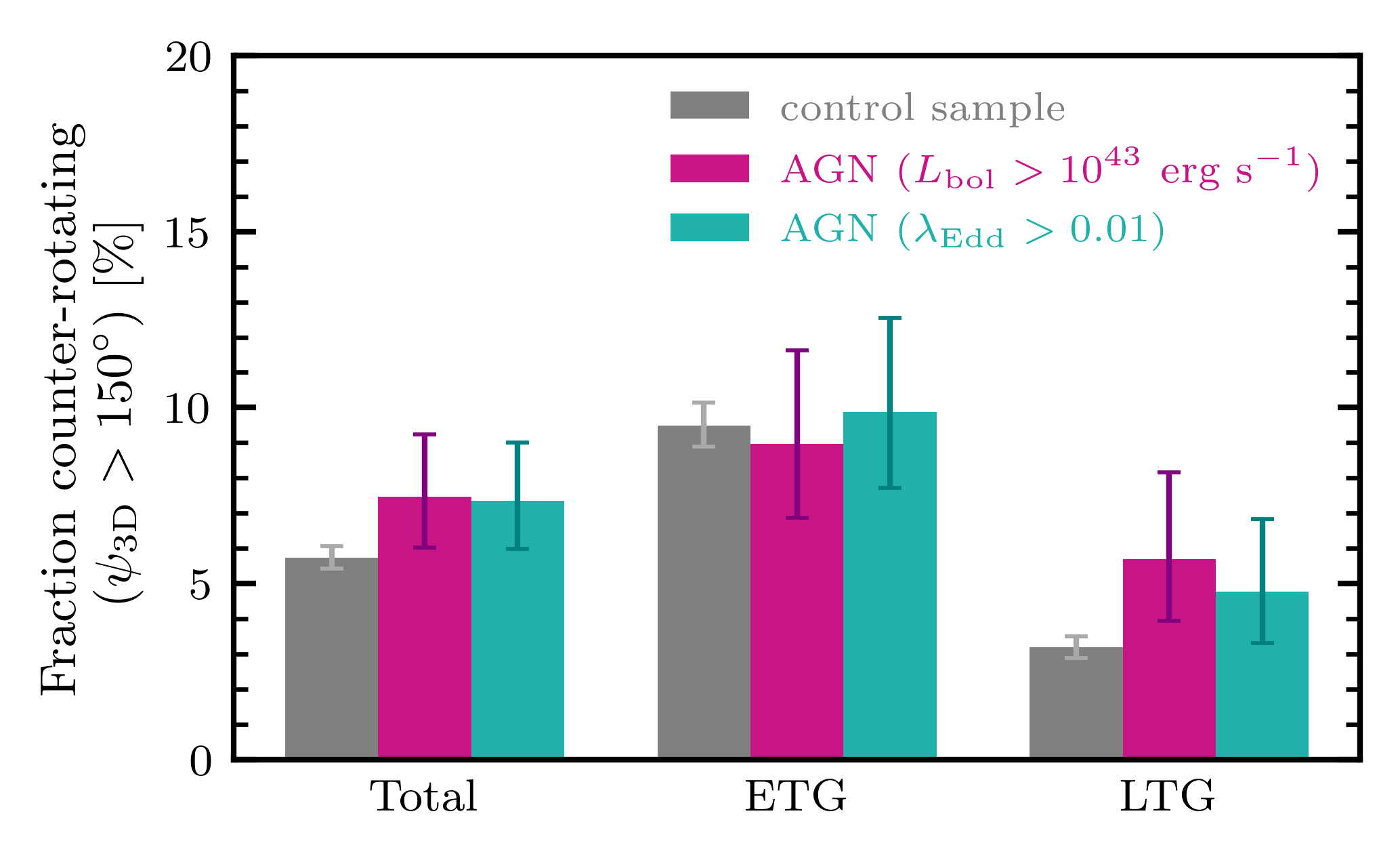}
    \caption{Fraction of misaligned (top) and counter-rotating (bottom) galaxies among the control sample (non-AGN plus AGN) at $z=0.1$ (grey, first bar of each group), and among two samples of AGN defined using $L_{\rm{bol}}>10^{43}$~erg~s$^{-1}$ (purple, second bar of each group), and $\lambda_{\mathrm{Edd}}>0.01$ (cyan, third bar of each group). Additionally, we show misalignment fractions among sub-samples of ETGs and LTGs. The error bars indicate the $1\sigma$ ($68$~per~cent) binomial confidence intervals estimated using the Jeffreys prior. We find a significantly higher fraction of misaligned galaxies in both AGN samples when compared to the control samples. We find only a significantly increased fraction of counter-rotating galaxies among the total and LTG samples for $L_{\rm{bol}}$-defined AGN, compared to their respective control samples.}
    \label{plot: f state AGN}
\end{figure}

In Figure~\ref{plot: f state AGN} we show the misaligned (top) and counter-rotating (bottom) galaxy fraction among the control sample (consisting of AGN and non-AGN galaxies) and both AGN samples at $z=0.1$. In total, misalignment fractions are significantly increased among both the $L_{\rm{bol}}$-defined AGN ($38.4\pm3.0$~per~cent) and the $\lambda_{\mathrm{Edd}}$-defined AGN ($33.8_{-2.7}^{+2.8}$~per~cent) compared to the control sample ($22.4\pm{0.6}$~per~cent). This increase is also seen individually among ETG and LTG populations. For ETGs, misalignment fractions of $58.6_{-4.1}^{+4.0}$~per~cent are found among $L_{\rm{bol}}$-defined AGN, compared to $45.0\pm1.1$~per~cent among the total ETG population \citep[see also][]{Baker2024}. For LTGs, misalignment fractions among $L_{\rm{bol}}$-defined AGN ($14.6_{-2.9}^{+3.5}$~per~cent) are almost double that of the typical LTG population $7.2\pm0.5$~per~cent. These results are all significant at the $2\sigma$ confidence level.

These results suggest that while unstable misalignments in LTGs can increase the likelihood of AGN activity, overall, other processes dominate AGN activity in LTGs. The most likely explanation for this is the relative rarity and short duration ($\sim0.4$~Gyr) of unstable misalignments in LTGs \citep[see ][]{Baker2024}. Consequently, many of these unstable misalignments in LTGs are not included as part of our analysis as they do not meet our window selection criteria. Furthermore, we note that our strict morphological classification excludes some systems that are morphologically transforming during these windows and which may be classified differently using instantaneous values of $\kappa_{\rm{co}}^*$.

We note that in all cases, the fraction of \textit{unstable} misaligned galaxies dominates the misaligned galaxy fraction. This is evident by the significantly lower counter-rotating fractions found among all galaxies when compared to their respective misalignment fractions. Indeed, the slightly increased counter-rotating fractions are only significant for $L_{\rm{bol}}$-defined AGN among the total and LTG populations. This is consistent with our results in Sections~\ref{results: enhanced BH growth total}~and~\ref{results: enhanced BH growth ETGs and LTGs} that showed BH growth is boosted only in systems with unstable misalignments. Furthermore, the weaker increase in counter-rotating fractions seen among our AGN samples is consistent with our results from Section~\ref{results: BH growth history at z=0.1}.

We conclude that the presence of an unstable misalignment increases the likelihood of AGN signatures at low redshifts in \textsc{eagle}. While a more detailed comparison of these results to observations is beyond the scope of this paper, these results are consistent with the observation that AGN signatures are more likely to be observed in misaligned systems \citep[e.g.][]{Raimundo2023, Raimundo2025} and that AGN luminosities are not boosted in counter-rotating systems \citep{Winiarska2025}. This is despite different morphological thresholds, AGN classifications, and misalignment uncertainties compared to observations.


\bsp	
\label{lastpage}
\end{document}